%% file: BSSBookch4.tex
\documentclass[graybox, envcountchap]{svmult}

\usepackage{mathptmx}        % selects Times Roman as basic font
\usepackage{helvet}          % selects Helvetica as sans-serif font
\usepackage{courier}         % selects Courier as typewriter font
\usepackage{makeidx}         % allows index generation
\usepackage{graphicx}        % standard LaTeX graphics tool
\usepackage{multicol}        % used for the two-column index
\usepackage[bottom]{footmisc}% places footnotes at page bottom

\makeindex             % used for the subject index
\input{common2}

\begin{document}
\pagestyle{empty}
\frontmatter%%%%%%%%%%%%%%%%%%%%%%%%%%%%%%%%%%%%%%%%%%%%%%%%%%%%%%

%\tableofcontents
%\include{cblist}
%\include{acronym}

\mainmatter%%%%%%%%%%%%%%%%%%%%%%%%%%%%%%%%%%%%%%%%%%%%%%%%%%%%%%%

\include{preston}
\backmatter%%%%%%%%%%%%%%%%%%%%%%%%%%%%%%%%%%%%%%%%%%%%%%%%%%%%%%%
%\appendix
%\include{appendix}
%\include{glossary}
\printindex

%%%%%%%%%%%%%%%%%%%%%%%%%%%%%%%%%%%%%%%%%%%%%%%%%%%%%%%%%%%%%%%%%%%%%%

\end{document}

%% file: common2.tex
\usepackage{amsmath}
\usepackage{amssymb}
\usepackage{lscape}
\usepackage{multirow}

\def\gapp{\ifmmode\stackrel{>}{_{\sim}}\else$\stackrel{<}{_{\sim}}$\fi}
\def\gsim{\lower.5ex\hbox{\gtsima}}
\def\gtsima{$\; \buildrel > \over \sim \;$}
\def\lapp{\ifmmode\stackrel{<}{_{\sim}}\else$\stackrel{<}{_{\sim}}$\fi}
\def\lsim{\lower.5ex\hbox{\ltsima}}
\def\ltsima{$\; \buildrel < \over \sim \;$}

\newcommand\apgt{\ {\raise-.5ex\hbox{$\buildrel>\over\sim$}}\ }
\newcommand\aplt{\ {\raise-.5ex\hbox{$\buildrel<\over\sim$}}\ }
\newcommand{\arcsec}     {\mbox{\ensuremath{{}^{\prime\prime}}}}%

%% file: preston.tex
\setcounter{chapter}{3}

\title{Field Blue Stragglers and Related Mass Transfer Issues}
\author{George W. Preston}
\institute{George W. Preston \at Carnegie Observatories, 813 Sana Barbara Street, Pasadena, CA 91101, USA\\ \email{gwp@obs.carnegiescience.edu}}
%\and Name of Second Author \at Name, Address of Institute \email{name@email.address}}
\maketitle
\label{Chapter:Preston}

\abstract*{This chapter contains my 
impressions and perspectives about the current state of knowledge about field blue stragglers (FBS) stars, drawn from an 
extensive literature that I searched. 
I conclude my review of issues that attend FBS and mass transfer, by a brief enumeration of a few mildly disquieting observational facts. }

\section{Introduction}
\label{presec:1}
I view field blue stragglers (FBS) through the prism of an aging observer.  This chapter contains my impressions and perspectives about the current state 
of knowledge about FBSSs, drawn from an extensive literature that I searched with the aid of the \emph{NASA Astrophysics Data System}\footnote{http://adsabs.harvard.edu/abstract\_service.html}, and \emph{Google}. The search prior to 2000 is spotty and I make no claim of completeness.   
In documenting various topics I tried to use references that acknowledge important earlier investigations. 

It is not possible to construct a sensible story about blue stragglers in the Galactic field\index{galactic field} without reference to their occurrence in 
globular clusters where they were first identified, or in Milky Way\index{Milky Way} satellites where they are now being discovered.  I have tried to 
minimise discussion of these topics, which are treated at length elsewhere in this book.  I include discussions of theoretical work 
(with diffidence) only when such work has an immediate impact on the interpretation of observations.

\subsection{Historical developments in first part of the $20^{th}$ century}
\label{presubsec:1.1}

My point of view in this presentation is that McCrea's \cite{mc64} mass transfer\index{mass transfer} produces the bulk of metal-poor\index{metal-poor} FBS in the Galactic halo\index{galactic halo}, so 
my history is largely confined to this process.  The earliest reference to ``mass transfer'' generated by my search engines is that of Kuiper \cite{ku41}, who 
introduced Roche lobe overflow\index{Roche lobe overflow} (RLOF) in his analysis of $\beta$ Lyrae\index{$\beta$ Lyrae}.  Kuiper \cite{ku41}, with characteristic modesty, refers to his own 
unpublished work conducted prior to his PhD thesis (1932) thusly: 
\begin{quotation}
..., a stream of matter from $M_A$ to $M_B$ is likely to set in...
\end{quotation}
And later,
\begin{quotation}
... the degree of contact increases very considerably during the process of {\bf mass transfer}...
\end{quotation}
In the 1930--40's Otto Struve assembled a stable of luminaries at Yerkes Observatory (Chandrasekhar, Greenstein, Henyey, Herzberg, Kuiper, Morgan, 
M\"unchen, Str\"omgren).  Though Kuiper had already established himself as an authority in binary stars, Struve \cite{st41} directed the attack on $\beta$ 
Lyrae by Kuiper and his colleagues at Yerkes (see \cite{sa78} for details).  For Struve $\beta$ Lyrae was an obsession.  He once confided 
to me that his goal (sadly, never-to-be-realised) was to spend his retirement at a modest telescope in the backyard of his home in the Berkeley hills, 
observing the endless spectroscopic variations produced by the gas-stream escaping from the outer Lagrangian point\index{Lagrangian point} $L_2$. 

Following WWII, Kopal \cite{ko55} invented the definitive binary classification scheme that persists to this day (detached, semi-detached, common envelope --- see Chap. 7),
and in 1967 groups led by Kippenhahn (G\"ottingen), Paczynski (Warsaw), and Plavec (Ondrejov) began detailed exploration of the consequences of
stellar evolution within the confines of a Roche potential. Kippenhahn  \& Weigert \cite{ki67} gave us cases A, B, and C, and Paczynski\cite{pa67} is generally
credited with resolving the Algol paradox\index{Algol paradox}, for which Plavec \cite{pa67} gave this gracious tribute: 
\begin{quotation}
May I say that until today it was the theory of close
binaries that lagged behind the observations.  Today we have a historical moment when the problem of the existence of semi-detached systems has been 
essentially solved theoretically.\end{quotation}

McCrea \cite{mc64} worked largely aloof from developments in continental Europe. His seminal paper, directed specifically to the ``extended main sequence of
some stellar clusters'', made no reference to Sandage's paper \cite{sa53} nor to any of the research referred to in the paragraph above, and it was his only publication
about binary stars.  However, I know that he had an abiding interest in binary stars, because a decade earlier I, then a graduate student, attended a 
number of his lectures devoted to problems of interacting binaries during his tenure as a Visiting Professor at Berkeley in 1956.  McCrea was invited to 
Berkeley by (\emph{who else?}) my mentor Otto Struve!

Speculations about the origin of blue stragglers in the halo field are inextricably linked to ideas about the origin of the Galactic halo itself.  
In addition to an origin {\it in situ}, via ELS \cite{eg62} and its various possible aftermaths, two additional hypotheses have 
been advanced. I \cite{pre94a} suggested that a modest fraction of them arise from capture of Galactic satellites similar to the present-day 
Carina dSph\index{dwarf spheroidal}, where they abound as part of an intermediate-age population \cite{sm94} or an even younger ``blue-plume'' 
\cite{mo04}. Gnedin \& Ostriker \cite{gn97}, on the other hand, calculated that the destruction time-scale for Galactic globular clusters\index{globular cluster} (hereafter GCs) 
is comparable to their lifetimes, and therefore that remnants of now-largely-destroyed primordial clusters might constitute a substantial fraction of the
 spheroid (bulge/halo) stellar population.  I shall return to these issues in Section~\ref{presec:5} below.

\section{Identification of BSS}
\label{presec:2}

\subsection{The metal-poor halo}
\label{presubsec:2.1}

The observational requirements for selection of (apparently young) halo FBSSs are parameters that (a) identify stars with the metallicity of the parent 
population and (b) sort them unambiguously by age.  The first of these two parameters must cull halo stars from the much more numerous, metal-rich, and 
age-heterogeneous populations of the thick and thin discs.  Note that there are complications here because of the overlapping abundance distributions of 
the three principal Galactic populations.  The second parameter must separate the ``young'' FBSSs from the bulk of their older main sequence cousins.  
If we ignore well-known details about the behavior of O and Na \cite{ca09}, and the poorly-constrained and indirectly inferred He content 
\cite{bu83}, the gross chemical compositions\index{chemical composition} of most GCs are homogeneous, and their ages can be estimated by fits of isochrones\index{isochrone} (there is a 
plethora of these in the literature) to the stellar distributions in colour-magnitude diagrams (CMD).  The fits are characterised most sensitively by colours of the 
main sequence turnoffs (MSTO).  Because of the great ages of GCs the variation of turnoff\index{turnoff} colour with age is small, $d(B-V)/dt\sim0.008$ dexGyr$^{-1}$ 
\cite{yi01}.  For this reason a number of small and not-so-small effects mask the age effect: interstellar (IS) reddening\index{interstellar reddening} , chemical composition, 
including [Fe/H]\index{metallicity}, and [alpha/Fe], and (assumed) He, the choice of mixing-length parameter, and the effects of radiation pressure and gravitational settling 
on stellar models. These effects combine to produce a blurred lower bound for temperature-sensitive colour indices of halo main sequence stars.

\subsubsection{Photometric criteria: UBV}
\label{presubsec:2.1.1}
%%%%%%%%%%%%%%%%%%%%%%%%%%%%%%%%%%%%%%%%%% FIGURE 1
\begin{figure}
\sidecaption
\includegraphics[width=75mm]{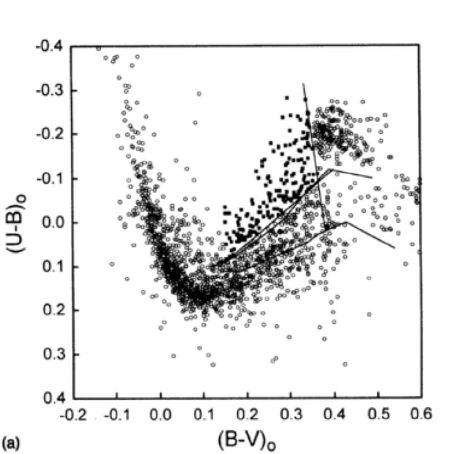}
%
% If no graphics program available, insert a blank space i.e. use
%\picplace{5cm}{2cm} % Give the correct figure height and width in cm
%
\caption{FBSSs in the $UBV$ two-colour plane lie in a triangular region blueward of a steeply-inclined blanketing vector that extends upward from the 
intrinsic two colour-relation for solar-type stars near $(B-V)_o=0.38$, and above a linearised MS relation for [Fe/H] $= -1$.  This figure is reproduced from 
\cite{pre94b} with permission from the Astronomical Journal.}
\label{prefig:1}       % Give a unique label
\end{figure}
%%%%%%%%%%%%%%%%%%%%%%%%%%%%%%%%%%%%%%%%%% FIGURE 1

The $UBV$ photometric\index{photometry} system identifies FBSSs \cite{pre94b} with reasonable success because $U-B$ is a sensitive metallicity indicator 
\cite{wa60,wi62,me60}, and $B-V$ is an adequate, though not ideal, temperature indicator 
(\cite{bu78, va03}).  The manner in which  the $UBV$ two-colour diagram isolates FBS candidates is indicated in Fig.~\ref{prefig:1} 
taken from \cite{pre94b}, which utilised photometry \cite{pre91} of the HK search for metal-poor stars~\cite{bee88}.  
 
\begin{figure}[b]
\sidecaption
\includegraphics[width=119mm]{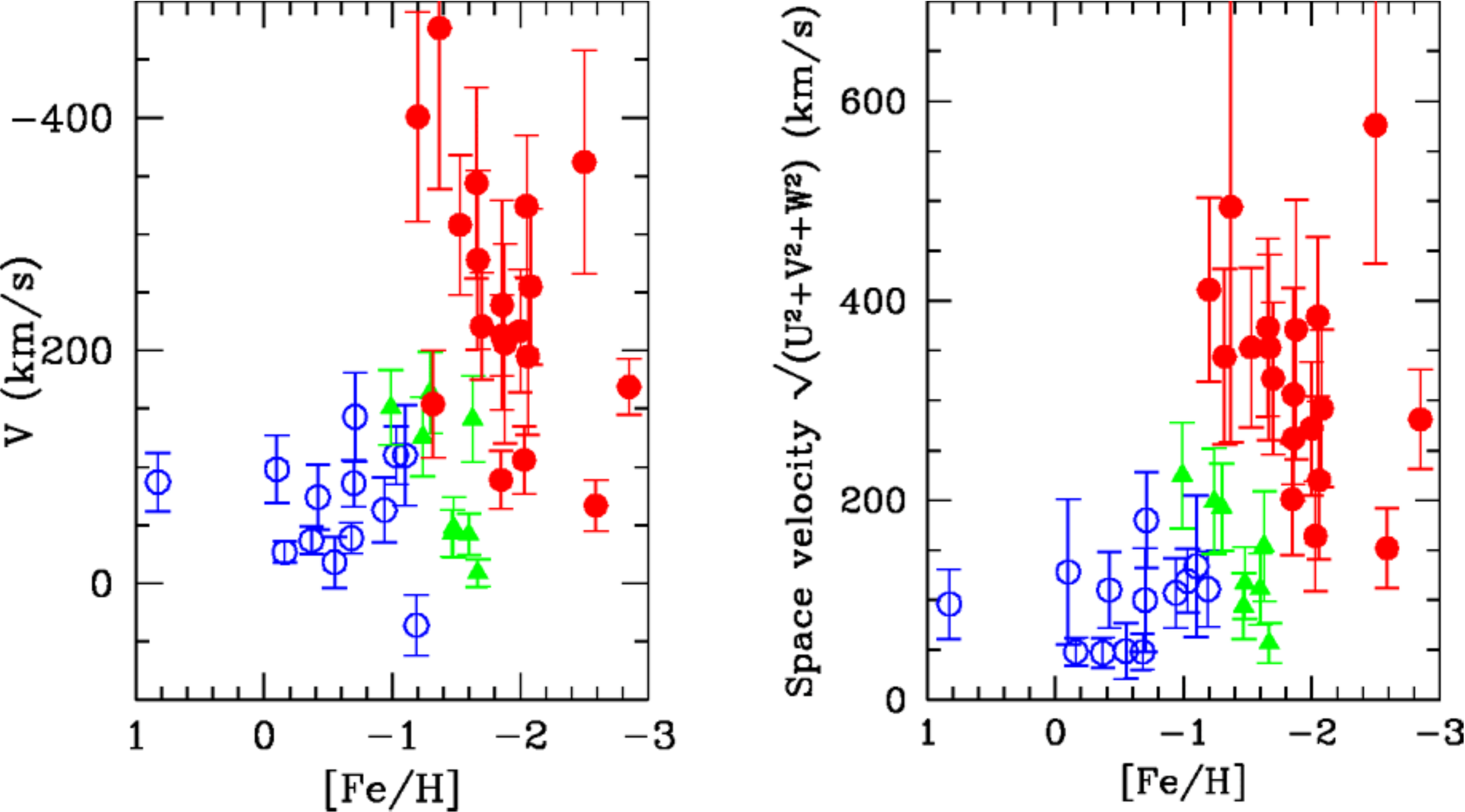}
%
% If no graphics program available, insert a blank space i.e. use
%\picplace{5cm}{2cm} % Give the correct figure height and width in cm
%
\caption{Galactic rotation vector (left panel) and space velocity (right panel) versus [Fe/H] for the 
PS00 sample of FBSSs.  The Figure is reproduced from \cite{br05} with permission from the IAU.}
\label{prefig:2}       % Give a unique label
\end{figure}

A steeply inclined blanketing vector extending upward from the intrinsic two-colour relation for solar-type stars \cite{wi62} near $(B-V)_o=0.38$ 
separates FBS candidates from the dense clump of metal-poor ``subwarfs'' near $(B-V)_o=0.4$, $(U-B)_o=-0.2$.  The UBV system\index{UBV system} fails below 
$(B-V)_o=0.15$ ($T_{eff} > 7700$ K), where line blanketing becomes too small to act as an effective abundance indicator.  Temperature-dependent 
blanketing vectors (\cite{wi62}) were used to construct a linear lower bound for [Fe/H]=$-1$.  The success of this lower bound can be judged by the 
plot of space motions versus [Fe/H] values derived from spectrum analysis by Preston \& Sneden (\cite{pre00},
hereafter PS00) of FBSSs in Fig.~\ref{prefig:2}, reproduced from~\cite{br05}.  

%%%%%%%%%%%%%%%%%%%%%%%%%%%%%%%%%%%%%%%%%% FIGURE 2
%%%%%%%%%%%%%%%%%%%%%%%%%%%%%%%%%%%%%%%%%% FIGURE 2

\begin{center}
\begin{figure}
\sidecaption
\includegraphics[width=75mm]{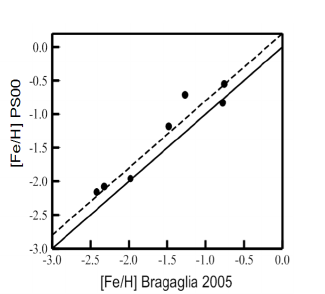}
%
% If no graphics program available, insert a blank space i.e. use
%\picplace{5cm}{2cm} % Give the correct figure height and width in cm
%
\caption{[Fe/H] values of PS00 versus those of~\cite{br05}.}
\label{prefig:3}       % Give a unique label
\end{figure}
\end{center}

Fig.~\ref{prefig:2} exhibits at once the strength and weakness of the $U-B$ abundance discriminant.  A clear majority of the stars (the red symbols and most of the green 
symbols) have spaces velocities and abundances ([Fe/H]$< -1.2$) associated with the halo.  However, the lower bound failed to eliminate a significant 
number of stars with [Fe/H]$>-1.0$.  This failure can be attributed in part at least to photometric errors and the vertical density gradient 
$dn/d(U-B)$ near the lower bound in Fig.~\ref{prefig:1} which combine to provide net spillover of metal-rich stars into the FBS  domain.  New abundances for stars in 
the PS00 survey derived from spectra with superior signal-to-noise ratios (S/N) and resolution \cite{br05} are plotted against those of PS00 in 
Fig.~\ref{prefig:3}.  The solid 45-degree line indicates 1:1 correspondence.  

%%%%%%%%%%%%%%%%%%%%%%%%%%%%%%%%%%%%%%%%%% FIGURE 3
%%%%%%%%%%%%%%%%%%%%%%%%%%%%%%%%%%%%%%%%%% FIGURE 3

Bragaglia et al \cite{br05} obtain abundances systematically lower than PS00 by $\sim0.15$ dex. Scatter in this regression contributes to blurring of the lower 
boundary of the FBS domain.

There is a bright side to this otherwise dreary discussion.  The binary fraction of stars in the PS00 survey with thick disc abundances ([Fe/H]$>-1$) is 
marginally higher, 0.76, than the fraction, 0.69, for stars with [Fe/H]$<-1$ and markedly higher than the spectroscopic binary fractions, $\sim0.15$, of the 
normal stars of the halo (\cite{la02}) and disc near the sun \cite{du91}.  Evidently, mild $(U-B)$ excesses properly select mildly 
metal-deficient FBSSs in the thick disc.  This result suggests that the bulk of the thick disc stars have similar ages, i.e., the thick disc does not contain a 
significant component of normal A-type and early F-type stars.  Most of the known FBSSs near the solar circle have been identified by $UBV$ photometry. 
For additions to the list of FBS candidates at distances less than about ~ 2 kpc subsequent to \cite{pre94b}, see the compilation 
by Wilhelm et al. \cite{wi99a}.

\subsubsection{Photometric criteria: SDSS\index{SDSS} $ugr$}
\label{presubsec:2.1.2}

Numerous FBS candidates have been identified in the distant halo by use of the ($u-g, g-r$) equivalent of the ($U-B, B-V$) diagram shown in 
Fig.~\ref{prefig:1} (\cite{ya00,si04}).  Noting that the conversion of ($u-g$) to ($U-B$) is not well-defined, Yanny et al. \cite{ya00} tentatively identify 
2715 FBSSs and 1493 BHB stars. Sirko et al. \cite{si04}, finding that $ugr$ photometry did not separate BHB and FBS satisfactorily, preferred the
use of Balmer lines as described in section~\ref{presubsec:2.1.5} below.  

FBSSs enter most discussions of BHB stars as a nuisance population that complicates the calculation of space densities of BHB stars.  
I have found little discussion of the apparent FBS population (e.g., space densities, Galactic distribution, clumping, specific frequency, 
abundance distribution) in the SDSS literature.  More about this possibility in section~\ref{presubsec:4.1} below.

\subsubsection{Other photometric criteria}
\label{presubsec:2.1.3}

Morrison et al. \cite{mo00} used the Washington photometric system\index{Washington photometric system} to search for halo substructure produced by satellite accretion.  
They chose BMP stars (\cite{pre94a}), of which FBSSs are a subset, as one of their halo tracers.  Using Washington $M-T_2$ as a 
temperature indicator they found 23 BMP to $V=19$ in an area of 2.75 deg$^2$ in accord with expectations (see their Fig. 2).  
In their discussion apparent magnitude, which limits the sample to heights far above the Galactic plane, isolates stars with low [Fe/H].

FBS candidates have been identified in the Str\"omgren photometric surveys of Olsen \cite{ol79,ol80} who lists 84 candidates in the summary tables of 
his two papers.  Stetson \cite{st91} identifies 15 possible metal-poor FBSSs in his Table 9.  I found it difficult to assess the reliability of these assignments 
mainly because of the confusing graphical displays of results.  It is my impression that Str\"omgren\index{Str\"omgren photometry} filters are not ideal tools for wholesale identification of
FBS.  Narrow bandwidths of the filters are a major drawback.

%%%%%%%%%%%%%%%%%%%%%%%%%%%%%%%%%%%%%%%%%% FIGURE 4
\begin{center}
\begin{figure}[h]
\sidecaption
\includegraphics[width=119mm]{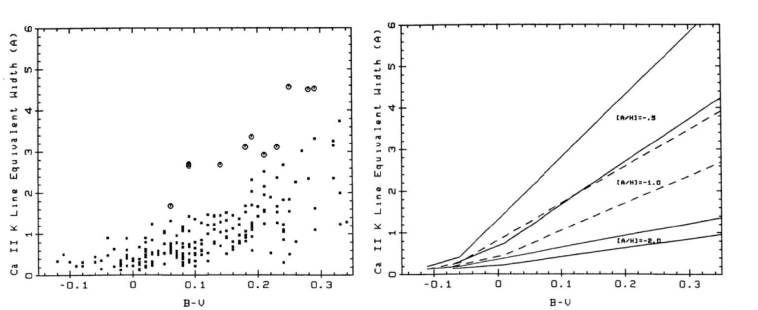}
%
% If no graphics program available, insert a blank space i.e. use
%\picplace{5cm}{2cm} % Give the correct figure height and width in cm
%
\caption{(left panel) Equivalent widths of CaII (K) for candidate BHB stars versus unreddened $B-V$ colour taken from \cite{pi83}; 
(right panel) abundance calibration of CaII (K) by \cite{ma78}. The figures are reproduced with permission of the AAS.}
\label{prefig:4}       % Give a unique label
\end{figure}
\end{center}
%%%%%%%%%%%%%%%%%%%%%%%%%%%%%%%%%%%%%%%%%% FIGURE 4

%%%%%%%%%%%%%%%%%%%%%%%%%%%%%%%%%%%%%%%%%% FIGURE 5
\begin{center}
\begin{figure}[b]
\sidecaption
\includegraphics[width=75mm]{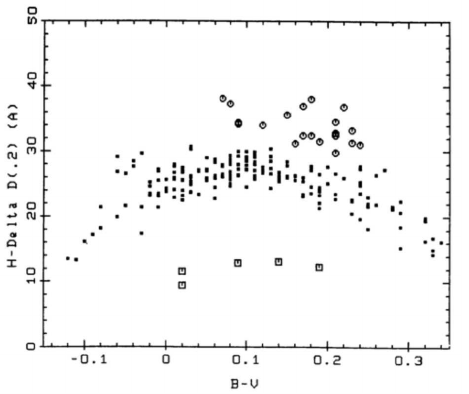}
%
% If no graphics program available, insert a blank space i.e. use
%\picplace{5cm}{2cm} % Give the correct figure height and width in cm
%
\caption{$D_{0.2}$, width of H$\delta$ at depth 0.2 below the local continuum, plotted versus unreddened $B-V$ for candidate BHB 
stars by \cite{pi83}.  Stars with main sequence gravities (open circles) are clearly separated from BHB stars (filled circles).  
The figure is reproduced with permission of the AAS.}
\label{prefig:5}       % Give a unique label
\end{figure}
\end{center}
%%%%%%%%%%%%%%%%%%%%%%%%%%%%%%%%%%%%%%%%%% FIGURE 5

\subsubsection{Hybrid spectro-photometric methods}
\label{presubsec:2.1.4}

Pier \cite{pi83} pioneered the identification of bona fide metal-poor blue horizontal branch stars (BHB) of the halo field and, thus indirectly, FBS 
stars by use of $B-V$ as a temperature coordinate, the equivalent width of the CaII K line as an abundance indicator, and the widths of 
Balmer lines as a gravity (luminosity) indicator. Following \cite{se66}, he estimated metal abundances by calibration of the 
CaII K line as shown by the diagrams in Fig.~\ref{prefig:4}, hereafter $BVK$ diagrams.

The left panel contains measured equivalent widths.  The right 
panel contains a calibration provided by \cite{ma78}.  Again 
following \cite{se66}, Pier \cite{pi83} showed that FBSSs and BHB stars could be disentangled in a 
HB candidate list by use of  $D_{0.2}$, the full width of a Balmer line (here H$\delta$) at a depth 20\% below the local continuum level.  
How well this works is illustrated in Fig.~\ref{prefig:5}. Here, as in several examples to follow, the FBS are treated as a nuisance to be removed from a BHB dataset.

Wilhelm et al. (\cite{wi99a}; method and \cite{wi99b}; results) compiled a list of 416 probable metal-poor FBSSs identified in the North and South
 hemisphere HK surveys (\cite{bee88,bee96}). Tools of the trade include $UBV$ photometry, the $BVK$ diagram\index{BVK diagram}, and $D_{0.2}$.

 %%%%%%%%%%%%%%%%%%%%%%%%%%%%%%%%%%%%%%%%%% FIGURE 6
\begin{figure}
\sidecaption
\includegraphics[width=119mm]{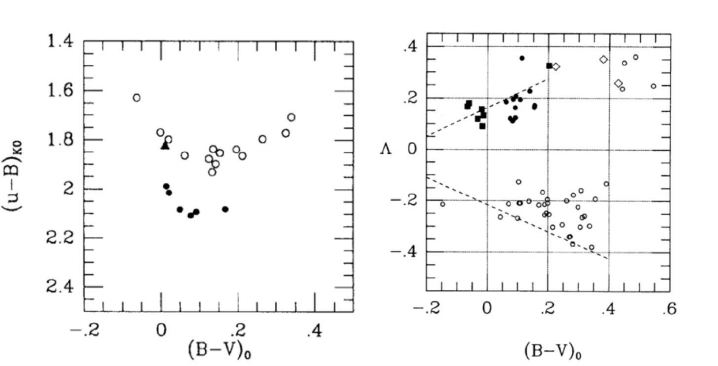}
%
% If no graphics program available, insert a blank space i.e. use
%\picplace{5cm}{2cm} % Give the correct figure height and width in cm
%
\caption{(left panel) The special 2-colour diagram of \cite{ki94} uses the Str\"omgren u filter to effect a clean separation 
of HB stars (filled circles) from main sequence stars (open circles); (right panel) measures steepness of the Balmer jump. The figures 
are reproduced with permission of the AAS.}
\label{prefig:6}       % Give a unique label
\end{figure}
%%%%%%%%%%%%%%%%%%%%%%%%%%%%%%%%%%%%%%%%%% FIGURE 6

%%%%%%%%%%%%%%%%%%%%%%%%%%%%%%%%%%%%%%%%%% FIGURE 7
\begin{center}
\begin{figure}[b]
\sidecaption
\includegraphics[width=75mm]{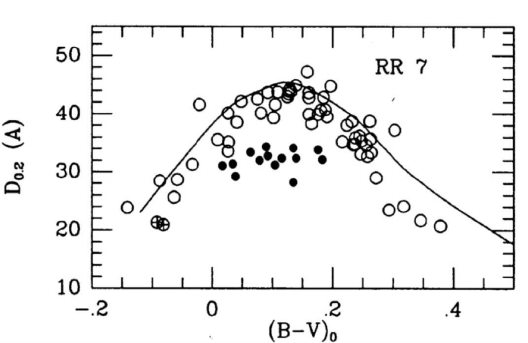}
%
% If no graphics program available, insert a blank space i.e. use
%\picplace{5cm}{2cm} % Give the correct figure height and width in cm
%
\caption{$D_{0.2}$, defined as the mean of the widths of H$\gamma$ and H$\delta$ at depth 0.2 below local continuum versus $(B-V)_o$.  Open and 
filled circles denote MS and HB stars, respectively.  This Figure is reproduced with permission of the AAS.}
\label{prefig:7}       % Give a unique label
\end{figure}
\end{center}
%%%%%%%%%%%%%%%%%%%%%%%%%%%%%%%%%%%%%%%%%% FIGURE 7

Kinman, Suntzeff \& Kraft \cite{ki94} use another hybrid method to separate BHB stars from FBSSs among AF stars of the Case Low-Dispersion 
 Northern Survey\index{Case Low-Dispersion Northern Survey} (\cite{sa88}).  Their classification criteria are: (1) a $(u-B)_K$ colour  (Str\"omgren $u$ filter and Johnson $B$ filter) that 
 measures the size of the Balmer jump, (2) a spectro-photometric index $\lambda$ that measures steepness of the Balmer jump, and (3) 
 the parameter $D_{0.2}$. The success with which these parameters separate BHB stars and FBSSs is shown in Fig.~\ref{prefig:6} and ~\ref{prefig:7}.

Concordance of classification assignments based on these three parameters is reassuring. Kinman et al. \cite{ki94} found that more than half of the A- and F-type stars 
fainter than $V=13.0$ and with $(B-V)_0<0.23$ are not BHB stars, but have surface gravities more like those expected for main-sequence stars.  
They remark that the more metal-poor of these are likely to be the blue metal-poor (BMP) stars discussed by \cite{pre94a}.

Flynn, Sommer-Larsen \& Christensen \cite{fl94} compared the performance of Str\"omgren $c1, (b-y)$  and $D_{0.2}$, $(B-V)$ photometry to separate BHB from FBS .  
They concluded that both systems work but that the $D(0.2), B-V$ system made more efficient use of telescope time. 

Finally, Carney et al. \cite{ca94} found a few ($n=6$) FBSSs in their radial velocity survey of high proper motion stars.  $(B-V)$ is their temperature parameter. 
They derived abundances by matching their spectra to a grid of synthetic spectra.  The low proper motion limits,  0.2\arcsec$yr^{-1}$, of the Giclas and Luyten 
surveys from which their sample was drawn confines the Carney et al. \cite{ca94} survey to a small volume of space in which only a few FBSSs are expected.

\subsubsection{Spectroscopic Criteria}
\label{presubsec:2.1.5}

%%%%%%%%%%%%%%%%%%%%%%%%%%%%%%%%%%%%%%%%%% FIGURE 8
\begin{center}
\begin{figure}[h]
\sidecaption
\includegraphics[width=119mm]{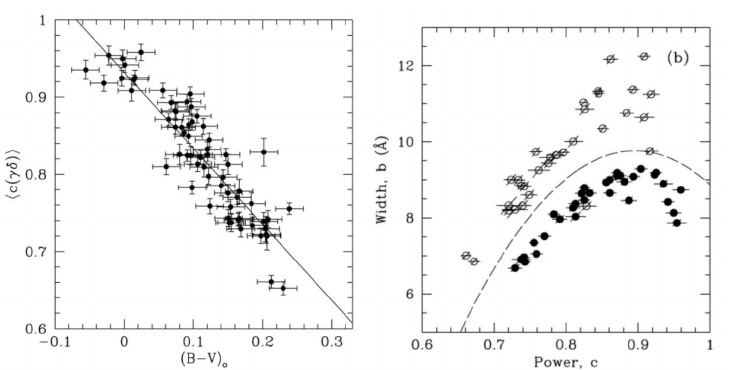}
%
% If no graphics program available, insert a blank space i.e. use
%\picplace{5cm}{2cm} % Give the correct figure height and width in cm
%
\caption{(left panel) A plot of the S\'ersic exponent $c$ versus $(B-V)_o$ taken from \cite{cl02}; (right panel) the S\'ersic scaled-width 
parameter $b$, effects an excellent separation of FBSSs (open circles) and BHB stars (filled circles).  The figure is reproduced with permission 
of Monthly Notices of the Royal Astronomical Society.}
\label{prefig:8}       % Give a unique label
\end{figure}
\end{center}
%%%%%%%%%%%%%%%%%%%%%%%%%%%%%%%%%%%%%%%%%% FIGURE 8

In a most refreshing example of cross fertilisation in astronomy Clewley et al. \cite{cl02} borrowed the S\'ersic \cite{se68} function from extragalactic astronomy 
to make analytical fits of observed Balmer line profiles in A-type stars.   They use the S\'ersic scale-width parameter $b$ and exponent $c$ to derive 
gravities and temperatures of A-type stars.   Fig.~\ref{prefig:8} shows that the S\'ersic exponent $c$ is well-correlated with $(B-V)_o$, hence, temperature.  
The scale-width parameter $b$ effects a clean gravity separation of metal-poor main sequence and BHB stars.  The $BVK$ diagram provides abundances.  
Brown et al \cite{bro05} compare the various hybrid methods for isolation of BHB stars from FBSSs (see their Fig. 5) from which it is evident that the 
S\'ersic parameters do the best job.

\begin{center}
\begin{figure}[h]
\sidecaption
\includegraphics[width=75mm]{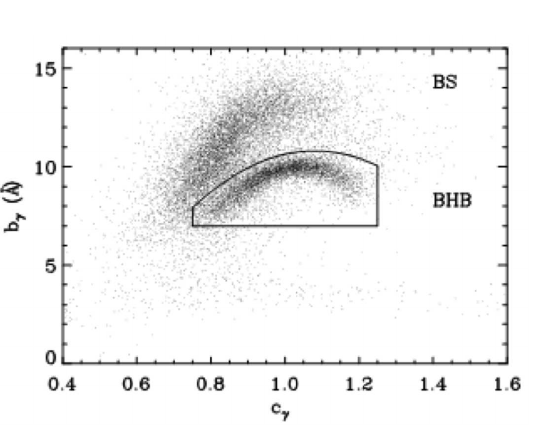}
%
% If no graphics program available, insert a blank space i.e. use
%\picplace{5cm}{2cm} % Give the correct figure height and width in cm
%
\caption{The S\'ersic parameters $b$ and $c$, employed by \cite{xu11}, effect a clear separation of FBS and BHB stars in 
the data of SDSS. The figure is reproduced with permission of the AAS.}
\label{prefig:9}       % Give a unique label
\end{figure}
\end{center}

Xue et al. \cite{xu11} used a similar application of the S\'ersic function\index{S\'ersic function} to effect a clean separation of FBSSs from BHB stars as shown in Fig.~\ref{prefig:9}, 
where the S\'ersic $b$ and $c$ parameters\index{S\'ersic parameter} were derived from measurements of H$\gamma$.  
Recall that $c$ is a temperature parameter, $b$ is a gravity parameter (the surface gravities of FBSSs exceed those of BHB stars by approximately one 
order-of-magnitude).  Because Xue et al. \cite{xu11} were primarily interested in kinematic substructure, they confined their analysis to the BHB stars. 
However, the FBS sample contains a wealth of additional untapped information about the stellar content of the halo.  

%%%%%%%%%%%%%%%%%%%%%%%%%%%%%%%%%%%%%%%%%% FIGURE 9
%%%%%%%%%%%%%%%%%%%%%%%%%%%%%%%%%%%%%%%%%% FIGURE 9

Fig.~\ref{prefig:9} contains about 
5000 BHB stars.  No count of FBSSs was given.  I guess conservatively by inspection that the number of FBSSs is comparable 
(recall Dustin Hoffman's count of toothpicks in the movie ``Rainman'').  An unknown fraction of the latter may be in as yet unidentified accretion streams.  

\begin{center}
\begin{figure}[h]
\sidecaption
\includegraphics[width=119mm]{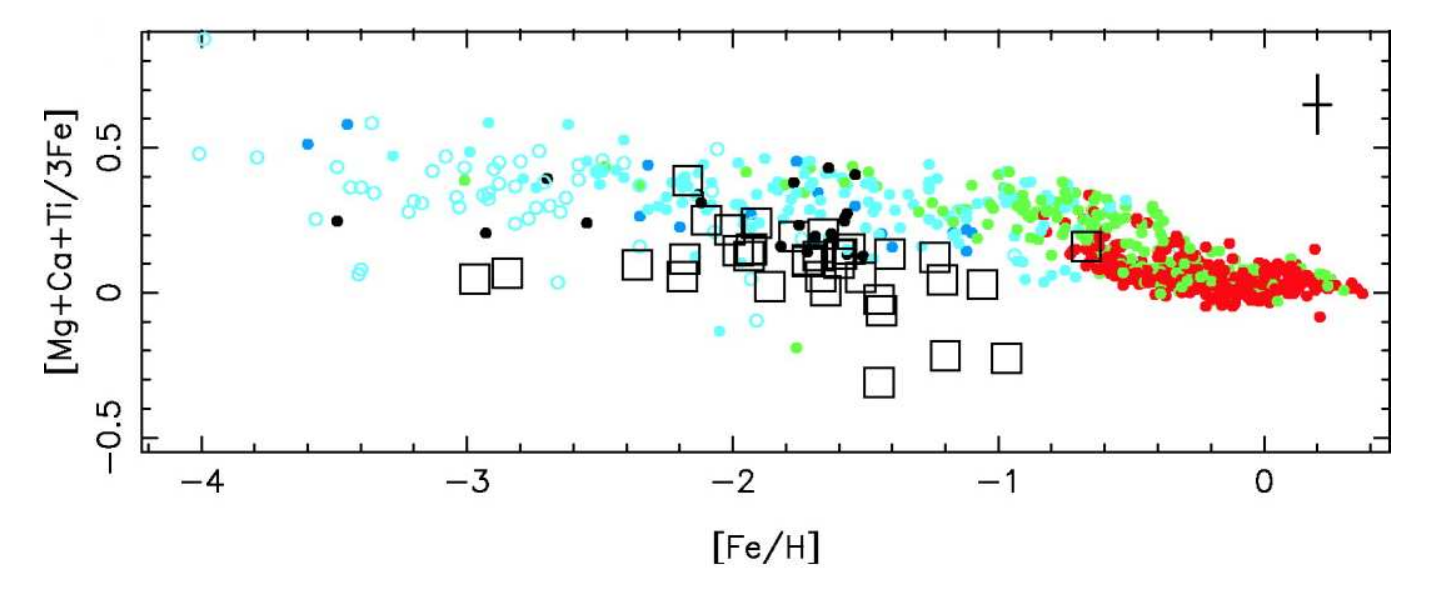}
%
% If no graphics program available, insert a blank space i.e. use
%\picplace{5cm}{2cm} % Give the correct figure height and width in cm
%
\caption{The [$\alpha$/Fe] values of red giants (boxes) in several dSph satellites of the Milky Way are systematically lower than the values 
for various species of Galactic stars (circles) in this compilation taken from Venn \cite{ve04}. The figure is reproduced with permission 
of the AAS.}
\label{prefig:10}       % Give a unique label
\end{figure}
\end{center}

This remarkable database can be used to resolve the Galactic halo BMP population into FBSSs formed in situ and FBSSs accreted from satellites.  
Recall that [$\alpha$/Fe] is systematically low by $\sim0.2$ dex at all [Fe/H] in the red giants of dwarf spheroidal satellites of the Milky Way relative to 
the bulk of Galactic halo stars.  This is illustrated in Fig.~\ref{prefig:10} taken from Venn \cite{ve04}, where black boxes denote red giants in the dwarf spheroidal galaxies\index{dwarf spheroidal galaxy} 
Carina, Draco, Fornax, Leo I, Sculptor, Sextans, and Ursa Minor \cite{ge05, sh01, sh03}, and the small circles denote 
various species of Galactic stars.  

%%%%%%%%%%%%%%%%%%%%%%%%%%%%%%%%%%%%%%%%%% FIGURE 10
%%%%%%%%%%%%%%%%%%%%%%%%%%%%%%%%%%%%%%%%%% FIGURE 10

Note further that two of the three $\alpha-$elements used to construct Fig.~\ref{prefig:10} possess prominent lines (Mg b, CaII K) that can be measured with 
modest spectral resolution, say $R=5000$.  Finally, note that an early exploration (PS00) yields promising results (see Section~\ref{presec:6} below).  
This is a project ready to go.

\subsection{FBS of the thick/thin disc}
\label{presubsec:2.2}

If McCrea's mass transfer produces most of the FBSSs, then we must be surrounded by lots of them in the thick/thin discs.   From an assemblage 
of star counts Sandage \cite{sa87} estimated stellar densities of the thin disc, thick disc, and halo to be in the proportions 200:22:1.  
Scaling the local metal-poor FBS number density, $\sim350$ kpc$^{-3}$, at the solar circle, by 100 in round numbers, we expect $\sim$35000 
disc FBSSs brighter than $V\sim$14.5, of which approximately 140 should be in the Henry Draper catalogue\index{Henry Draper catalogue}, and 10 in Olsen's surveys which 
were limited to $V<8.3$.  By this reckoning Olsen found too many disc FBSSs (see section~\ref{presubsec:2.1.3} above).  

A convincing procedure for wholesale identification of disc FBSSs, simpler in its application than detailed one-by-one 
analysis \cite{fu99,fu11}, has yet to be devised.  Perhaps, in view of the high binary fraction, 0.76, among 
PS00 sample stars with [Fe/H]$>-1$, we should revisit the $UBV$ identification criteria.

\section{Group Properties of Metal-Poor FBS}
\label{presec:3}

This section is devoted to a number of characteristics of metal-poor FBSSs that set them apart from other stellar groups.  
The first, to which all others are subsidiary, are the colour boundaries which define the group.  The boundaries are defined in a number 
of photometric systems\index{photometric system}.   

\subsection{Colour boundaries}
\label{presubsec:3.1}

\begin{center}
\begin{figure}[h]
\sidecaption
\includegraphics[width=119mm]{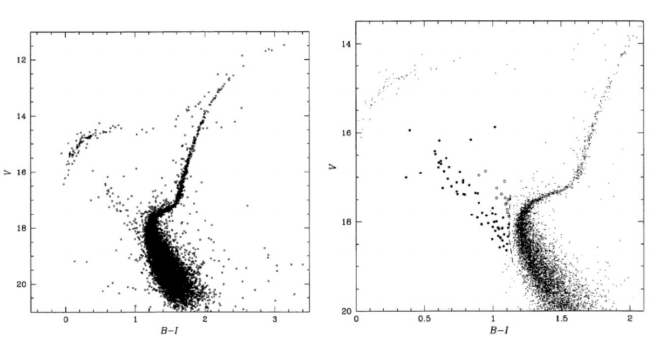}
%
% If no graphics program available, insert a blank space i.e. use
%\picplace{5cm}{2cm} % Give the correct figure height and width in cm
%
\caption{\emph{(left panel)} The photometry of \cite{ma97} clearly shows that a sequence of blue stragglers merges with the 
main sequence below the MSTO in the GC M55;  \emph{(right panel)}  The dashed line slightly blueward of the MSTO artificially 
truncates the sequence of mass-transfer binaries.  The figure is reproduced with permission of the AAS.}
\label{prefig:11}       % Give a unique label
\end{figure}
\end{center}

In all practical cases the red (cool) boundary of the blue straggler domain is defined to be at or near the colour of the MSTO of the stellar system in which 
they occur, i.e., the boundary of the group is defined by stars which are not members of the group.   Such a definition leads to the artificial edge in 
the beautiful photometry of NGC 6809\index{NGC 6809} \cite{ma97} indicated by the dashed curve in the right panel of Fig.~\ref{prefig:11}. 

%%%%%%%%%%%%%%%%%%%%%%%%%%%%%%%%%%%%%%%%%% FIGURE 11
%%%%%%%%%%%%%%%%%%%%%%%%%%%%%%%%%%%%%%%%%% FIGURE 11

An obvious (to me) sequence of mass transfer binaries extends continuously into the single-star main
sequence in the left panel.  Such a red extension was a natural ingredient of the PS00 toy model, and the more sophisticated applications 
of Eggleton's code \cite{eg73} by, for example, Tian et al. \cite{ti06} and Chen \& Han \cite{ch09} who, however, did not make calculations for old metal-poor 
systems considered here. Ryan et al. \cite{ry01, ry02} reach the same conclusion, arguing that Li-deficient binaries with greater-than-average axial 
rotation below the MSTO should be regarded as mass-transfer\index{mass transfer} binaries. 
 
A second logical defect of this definition is dramatised by considering the MSTO $(B-V)_o$ values for two well-studied clusters each of which contains BSS.  
The clusters M67\index{M67}, and NGC 5466\index{NGC 5466} have MSTO $(B-V)_o$ values of  0.50 \cite{va10} and 0.39~\cite{be09}, a large difference, 
comparable to the entire colour range of the FBSSs, due primarily to the $\sim2$ dex difference in metallicity.  FBSSs with this range in metallicity 
are common in the halo/disc field, where there are no MSTOs to guide us.  

What we realise from this cluster comparison is that there is no one-size-fits-all red (low temperature) boundary to the FBS domain.  Furthermore, the field 
contains sizable numbers of FBSSs with abundances well below the low limit, [Fe/H]$\sim-2.5$, of the globular cluster population, and a fraction of these are CEMP-s stars, formed by mass transfer from AGB\index{asymptotic giant branch star} companions, that pay no heed to the MSTO; outstanding, indubitable examples are CS 22964-161
\cite{th08} and CS 22949-008 \cite{mas12}.   An even larger fraction is formed by mass transfer from ordinary RGB\index{red giant} companions.  
As we shall see in Sect.~\ref{presec:6} below, knowledge of the amounts of mass transferred, whether by Roche lobe overflow\index{Roche lobe overflow} (RLOF) or wind accretion\index{wind accretion} is now or 
will soon become important for estimation of dilution of the accreted mass by the receiver envelope, for calculating changes in dilution modified by 
thermohaline mixing, for accurate calculation of the efficiency of wind accretion in the variety of scenarios posed by binaries with different orbital parameters, 
and for reconciliation of the relative numbers of MS and giant CEMP-s stars.

\begin{center}
\begin{figure}[h]
\sidecaption
\includegraphics[width=75mm]{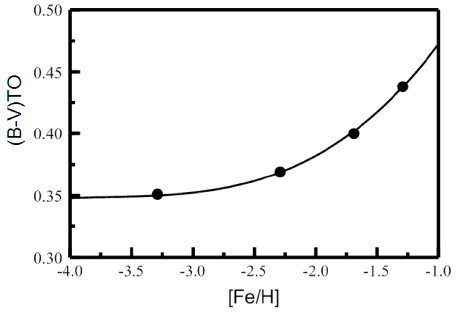}
%
% If no graphics program available, insert a blank space i.e. use
%\picplace{5cm}{2cm} % Give the correct figure height and width in cm
%
\caption{The $B-V$ colour of the MSTO versus [Fe/H] for an age of 13.5 Gyr interpolated among the isochrones of 
\cite{yi01} by use of a 3$^{rd}$ order polynomial.}
\label{prefig:12}       % Give a unique label
\end{figure}
\end{center}

It is possible to generalise the cool boundary for metal-poor FBSSs by converting it to a parameter that can be calculated for each star, rather than a 
parameter defined for the whole group.  Because of the large ages of metal-poor halo stars, MSTO colour varies only slowly with 
age, $d(B-V)/dt\sim0.008$ magGyr$^{-1}$ \cite{yi01}.  Therefore, with negligible error we may adopt a single, canonical age, say 13.5 Gyr, for 
the parent populations of all the individual metal-poor FBSSs.  The MSTO $B-V$ values of the Y$^2$ isochrones \cite{yi01} for age 13.5 Gyr 
are shown in Fig.~\ref{prefig:12}. 

%%%%%%%%%%%%%%%%%%%%%%%%%%%%%%%%%%%%%%%%%% FIGURE 12
%%%%%%%%%%%%%%%%%%%%%%%%%%%%%%%%%%%%%%%%%% FIGURE 12

[Fe/H] values and the $\Delta(B-V)$ colour displacements of each star from their respective MSTOs then yield {\it lower limits} to the accreted masses via  
$\Delta M/\Delta(B-V)$.  {\it Lower limits} are the best that can be achieved, because the initial masses of the receivers, hence their initial
$B-V$ colours\index{B-V colour}, are unknown:  they may lie within a considerable range of values at or below their MSTO values (PS00, \cite{ry02}).  
The [Fe/H] values of several FBSSs now under abundance scrutiny lie on $-3.5<$[Fe/H]$<-1.5$, which, from Fig.~\ref{prefig:12}, corresponds to a range in MSTO $B-V$ 
colour of $\sim0.08$ mag.  The individual displacements $\Delta(B-V)$ can be converted to accreted masses by values  $\Delta M/\Delta(B-V)$ derived 
from isochrones, typically $\sim-1.5$ M$_{\odot}$ mag$^{-1}$. I calculated accreted masses in this manner for several FBSSs of interest in Sect.~\ref{presec:6} below.

\subsection{Specific Frequencies}
\label{presubsec:3.2}

\begin{center}
\begin{figure}[h]
\sidecaption
\includegraphics[width=119mm]{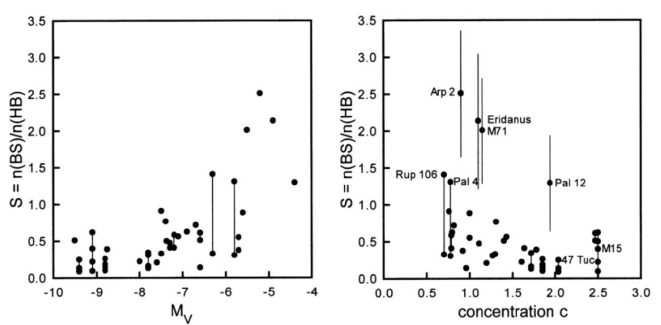}
%
% If no graphics program available, insert a blank space i.e. use
%\picplace{5cm}{2cm} % Give the correct figure height and width in cm
%
\caption{\emph{(left panel)} Specific frequency of BSSs versus GC absolute visual magnitude.  \emph{(right panel)}  Specific frequency of BS 
stars versus King concentration parameter.  The figures, taken from PS00, are reproduced with permission from the AAS.}
\label{prefig:13}       % Give a unique label
\end{figure}
\end{center}

To estimate the incidence of FBSSs in their parent population we \cite{pre94b} introduced specific frequencies, $S_{BMP}$, defined 
as the ratio of space densities of BMP and HB stars\index{horizontal branch} near the solar circle.  HB stars are a convenient reference population because both 
theory and observation support the notion that HB density is closely proportional to total population density.   
The HB models of \cite{le90} with typical halo abundance $Z=0.0004$ have lifetimes that range from $0.90\times10^8$ yr for mass
$M/M_{\odot}=0.52$ to $1.14\times10^8$ y for mass $M/M_{\odot}=0.76$, and the average number ratio, 1.38, of HB stars to red giants more 
luminous than the horizontal branch in metal-poor globular clusters \cite{bu83,zo00} has a standard deviation of only 0.15.  
Lower RGB stars with luminosities comparable to those of BSSs \cite{bo93} or MS stars near the turnoff \cite{ch09} 
are not options in the halo field, because such stars cannot be identified unambiguously. We \cite{pre94b} used our high value $S_{BMP}=8$ 
relative to the values $<1$ for BSSs in globular clusters to conclude that cluster-type BSSs are a minor constituent of the BMP population, and 
 suggested an extra-Galactic origin.  We were overly enthusiastic.  Subsequent recognition of the high binary fraction among BMP stars (PS00) 
produces a lower specific frequency for FBSSs, $S_{BMP}=4.0$, that is nevertheless an order-of-magnitude larger than those of the halo globular 
clusters, as shown in Fig.~\ref{prefig:13}.  Notice that $S_{BMP}$ values are conspicuously larger in the sparse globular clusters of the outer halo, identified by 
name in the right panel of Fig.~\ref{prefig:13}. 

%%%%%%%%%%%%%%%%%%%%%%%%%%%%%%%%%%%%%%%%%% FIGURE 13
%%%%%%%%%%%%%%%%%%%%%%%%%%%%%%%%%%%%%%%%%% FIGURE 13

\begin{center}
\begin{figure}[h]
\sidecaption
\includegraphics[width=119mm]{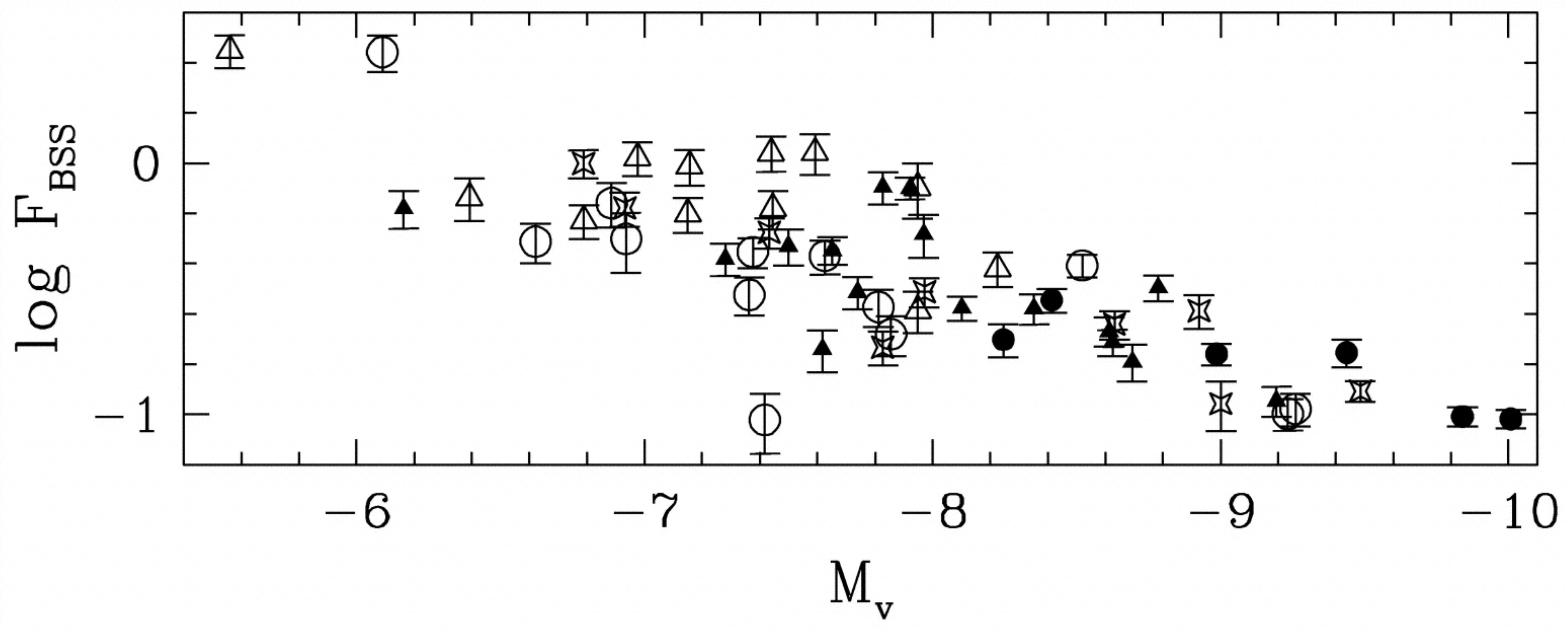}
%
% If no graphics program available, insert a blank space i.e. use
%\picplace{5cm}{2cm} % Give the correct figure height and width in cm
%
\caption{Logarithmic specific frequency of BSSs in globular clusters based on the HST\index{Hubble Space Telescope} data of \cite{pi04}.  
This figure is reproduced with permission from the AAS.}
\label{prefig:14}       % Give a unique label
\end{figure}
\end{center}

The marked decreases with increasing cluster Luminosity (mass) and the King \cite{ki66} concentration parameter $c$  played key roles in formulation of the 
view that collisional disruption of  long period (wide) binaries \cite{le89,ho01}, by far the most numerous in the orbital period 
distribution of Duquennoy \& Mayor \cite{du91}, accompanied by merger of the much less numerous short-period binaries (\cite{ma90}) are the two 
physical processes responsible for the low specific frequencies of BSSs in GCs relative to the field.  
Piotto et al. \cite{pi04} confirmed the correlation of specific frequency with
absolute luminosity and the relatively low specific frequencies with their elegant HST data for 56 GCs shown in Fig.~\ref{prefig:14}.  The upper bound 
to specific frequency among the oldest open clusters extended the correlation essentially to the field value of PS00 \cite{de06}.

%%%%%%%%%%%%%%%%%%%%%%%%%%%%%%%%%%%%%%%%%% FIGURE 14
%%%%%%%%%%%%%%%%%%%%%%%%%%%%%%%%%%%%%%%%%% FIGURE 14

The disruption of wide binaries in M22\index{M22} is contested by C\^ot\'e et al. \cite{co96}, {\it in spite of the fact that their own 22 yr search led to a marked deficiency of 
such binaries relative to the Galactic field}.  A related situation was encountered by Mayor, Duquennoy \& Udry \cite{may96} in $\omega$  Cen\index{$\omega$ Centauri}, who reported 
that only 2 of 32 CEMP stars in that cluster are spectroscopic binaries, contrary to the expectations of McClure \& Woodsworth \cite{mc90}.  At issue is whether 
CEMP stars in these two clusters are produced by McCrea-type star-by-star mass transfer or by pollution of intra-cluster gas during evolution of a previous 
generation of stars.  This topic has a fascinating history \cite{he77,no83,va94,ma09,ma12} beyond the purview of this chapter.

\subsection{The Distinguishing Characteristics of FBS Binary Orbits}
\label{presubsec:3.3}
The orbital characteristics of FBSSs, discussed at length by PS00 and Carney et al. \cite{ca01}, are these: 
\begin{description}
\item[(a)]  the mass functions of FBSSs are smaller by a factor of two than those for binaries of the disc \cite{du91} and high proper motion 
samples \cite{la02} as illustrated by the cumulative distributions in the left panel of Fig.~\ref{prefig:15}.  
McClure \& Woodsworth \cite{mc90} encountered a similar situation in their comparison of their gCH\index{CH giant} binaries with normal GK giant\index{red giant} binaries.  Comparisons 
of the cumulative distributions for these samples are reproduced in the right panel of Fig.~\ref{prefig:15}.  Both comparisons are consistent with the idea that 
FBS and gCH binaries have secondaries (white dwarfs) whose masses are systematically lower than the companion masses of their normal counterparts.

%%%%%%%%%%%%%%%%%%%%%%%%%%%%%%%%%%%%%%%%%% FIGURE 15
\begin{center}
\begin{figure}[h]
\sidecaption
\includegraphics[width=119mm]{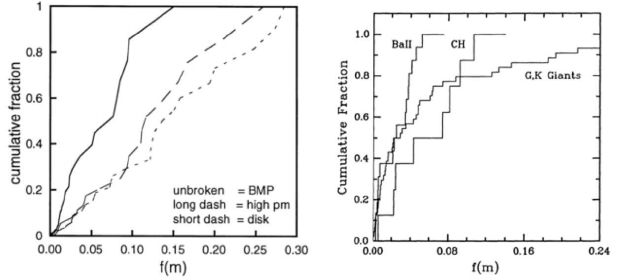}
%
% If no graphics program available, insert a blank space i.e. use
%\picplace{5cm}{2cm} % Give the correct figure height and width in cm
%
\caption{\emph{(left panel)} The cumulative distribution of mass functions for FBS binaries is compared to those for ordinary binaries of the disc and halo.  
The figure, taken from PS00, is reproduced with permission from the AAS.  \emph{(right panel)} The cumulative distributions of mass 
functions for Ba star\index{barium star} binaries and CH-giant binaries are compared to those for ordinary G-K giant binaries.  The figure, taken from 
\cite{mc90}, is reproduced with permission from the AAS.}
\label{prefig:15}       % Give a unique label
\end{figure}
\end{center}
%%%%%%%%%%%%%%%%%%%%%%%%%%%%%%%%%%%%%%%%%% FIGURE 15

\item[(b)] There is a deficit of short-period binaries and an excess of orbital eccentricities less than 0.2 among 
FBSSs (PS00) and CH stars (\cite{mc90,mc97}) compared to disc binaries (\cite{du91}) 
and high proper motion binaries (\cite{la02}).  Fig.~\ref{prefig:16} illustrates these 
comparisons of FBS binaries (middle panel) to their carbon-star cousins (bottom panel) and to the parent populations from which both are drawn (top panel).  
Straight-line segments join alternative orbital solutions for a number of FBS binaries obtained by PS00.  Such alias periods, three of them for three of the stars, 
were an unfortunate, inevitable consequence imposed by annual/biennial observing trips to Chile.  I included these alternatives in the middle panel of Fig.~\ref{prefig:16} to 
assure the reader that the conclusions about periods and eccentricities are not sensitive to which alternatives are chosen.  

%%%%%%%%%%%%%%%%%%%%%%%%%%%%%%%%%%%%%%%%%% FIGURE 16
\begin{center}
\begin{figure}[h]
\sidecaption
\includegraphics[width=75mm]{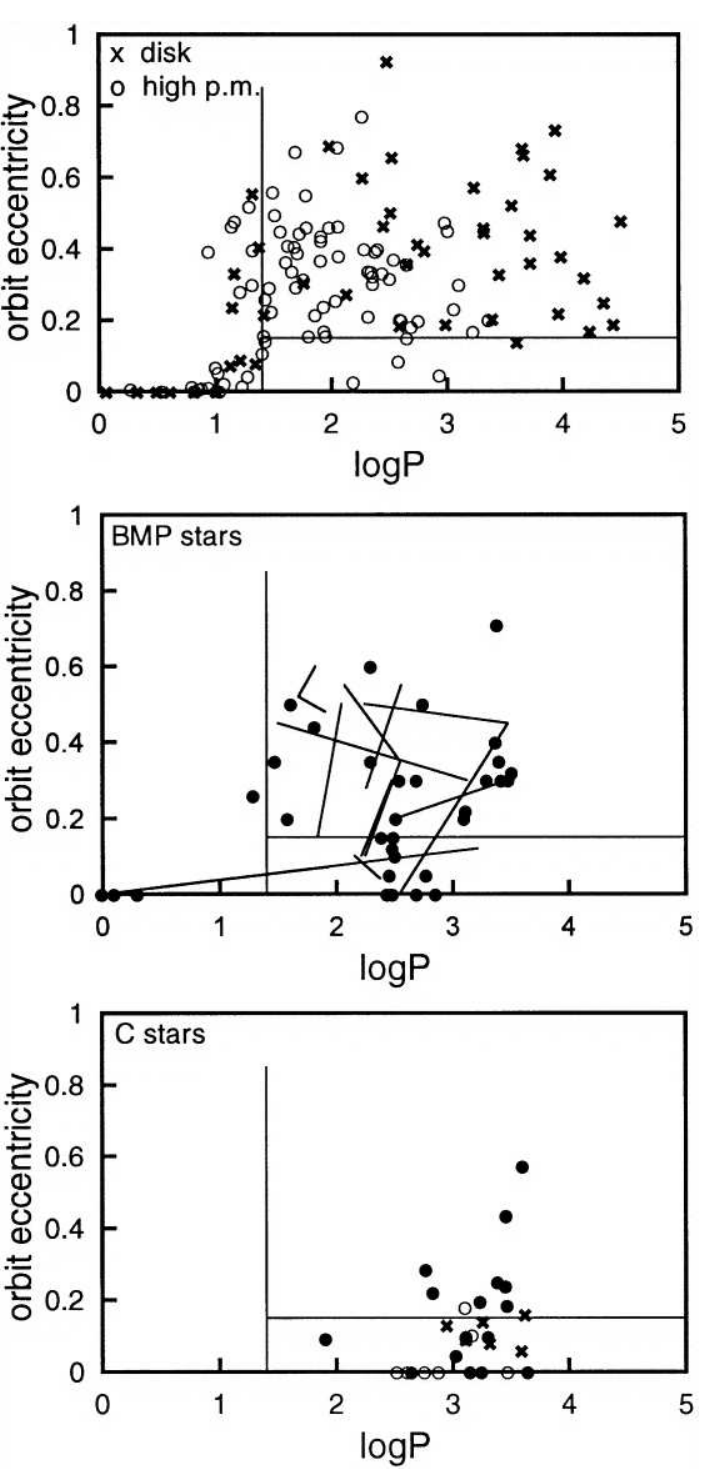}
%
% If no graphics program available, insert a blank space i.e. use
%\picplace{5cm}{2cm} % Give the correct figure height and width in cm
%
\caption{Orbital eccentricity versus logarithm of orbital period for (top panel) ordinary binaries of the disc (crosses) and high proper motion 
samples (open circles); (middle panel) FBSSs of PS00; (bottom panel) Ba stars (filled circles) and CH giants (open circles).  
The figures, taken from PS00, are reproduced with permission from the AAS.}
\label{prefig:16}       % Give a unique label
\end{figure}
\end{center}
%%%%%%%%%%%%%%%%%%%%%%%%%%%%%%%%%%%%%%%%%% FIGURE 16

It is commonly accepted that periods lengthen during conservative mass transfer after the mass of the donor falls below that of the receiver \cite{hi01} 
and that such mass transfer circularises orbits.  However, Sepinsky et al. \cite{se07, se09, se10} have revisited mass transfer with and without mass loss\index{mass loss} in 
binaries with eccentric orbits\index{eccentric orbit} and obtain a wide variety of outcomes that depend on initial conditions.  Even in the conservative case periods and 
eccentricities may either increase or decrease on time scales that range from a few million years to a Hubble time.  FBSSs and CEMP-s stars provide a 
rich database with which to test these recent theoretical developments.

\item[(c)]  Finally, PS00 found no double-lined spectroscopic binaries (SB2) among their FBSSs while SB2 comprise 20\% of the binaries in the high proper 
motion sample of Latham et al. \cite{la02}. The absence of detectable secondaries among FBS binaries is a natural consequence of the mass transfer hypothesis.  
I \cite{pre94a} did discover one extremely metal-poor SB2, CS 22873-139 and deduced an age of 8 Gyr from a discussion of the observed colours, 
suggesting that this binary came to the Milky Way from a satellite like the Carina dSph.  This result was challenged by Spite et al.\cite{sp00} who derived a lower 
$T_{eff}$ from the H$\alpha$ profile. Interestingly, however, they also found low [$\alpha$/Fe]$\sim0$ and did not detect lithium --- results expected for a 
BSS accreted from a dSph (see Fig.~\ref{prefig:10} and the discussion in section~\ref{presubsec:6.2}) 
\end{description}

To summarise: the orbital characteristics of the FBS binaries are distinctly different from the ordinary binaries of their parent populations with regard to
 mass functions\index{binary mass function}, orbital periods\index{orbital period}, orbital eccentricities, and incidence of detectable secondaries.  FBS binaries are a species {\it sui generis}. 

\section{Galactic Distribution}
\label{presec:4}

\subsection{Smooth Halo Field}
\label{presubsec:4.1}

\begin{center}
\begin{figure}[t]
\sidecaption
\includegraphics[width=119mm]{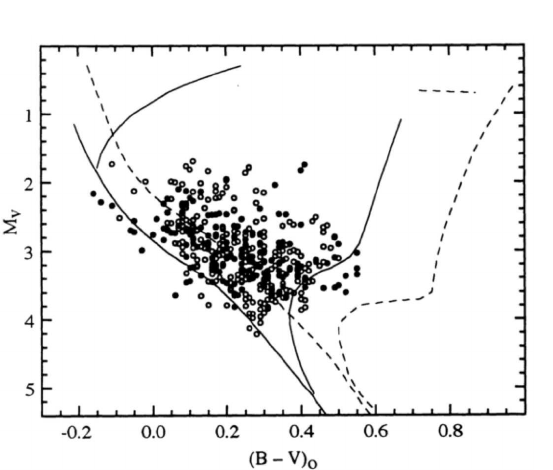}
%
% If no graphics program available, insert a blank space i.e. use
%\picplace{5cm}{2cm} % Give the correct figure height and width in cm
%
\caption{A composite colour-magnitude diagram from GC BSSs taken from \cite{sa93}.  This figure is reproduced with 
permission of the Publications of the Astronomical Society of the Pacific.}
\label{prefig:17}       % Give a unique label
\end{figure}
\end{center}

SDSS-based\index{SDSS} studies have identified thousands of FBSSs widely dispersed over the distant halo (see Fig \ref{prefig:4} and \ref{prefig:10} above).  
These FBSSs have been identified primarily for the purpose of isolating pure BHB samples; they have not been used, to my knowledge, 
in investigations of halo structure.  The compilation of BSSs in GC by Sarajedini \cite{sa93} displayed in Fig.~\ref{prefig:17} allows estimation of their utility for such purposes.  

%%%%%%%%%%%%%%%%%%%%%%%%%%%%%%%%%%%%%%%%%% FIGURE 17
%%%%%%%%%%%%%%%%%%%%%%%%%%%%%%%%%%%%%%%%%% FIGURE 17

For simplicity of presentation I did not use the much busier update of this diagram in \cite{co12}.  Supposing that the ranges of 
age and chemical composition in the GC sample 
are similar to those encountered in the halo, I estimate from inspection of Fig.~\ref{prefig:17} that the standard deviation of $M_V$ for FBSSs on $0.0<(B-V)_o<0.4$ 
is $\sigma < 0.5$ mag., which translates to 25\% errors in distances to individual stars.  The range of galacto-centric distances over which SDSS FBS 
data could be used to calculate space densities is $\sim20$ kpc, so my gloss on the situation is that FBSSs may be useful for investigations of space 
density on such length scales.  How useful they might be would require more careful error analysis.  Nevertheless, FBS have not been entirely 
discounted for such purposes. Univane, \& Gilmore \cite{un96} used star counts blue-ward of the classical halo turnoff at $B-V\sim0.4$ to estimate 
that $\sim10$ per cent of the halo could be attributed to accretion of intermediate-age satellites.

\subsection{The Galactic Bulge}
\label{presubsec:4.2}

\begin{center}
\begin{figure}[h]
\sidecaption
\includegraphics[width=119mm]{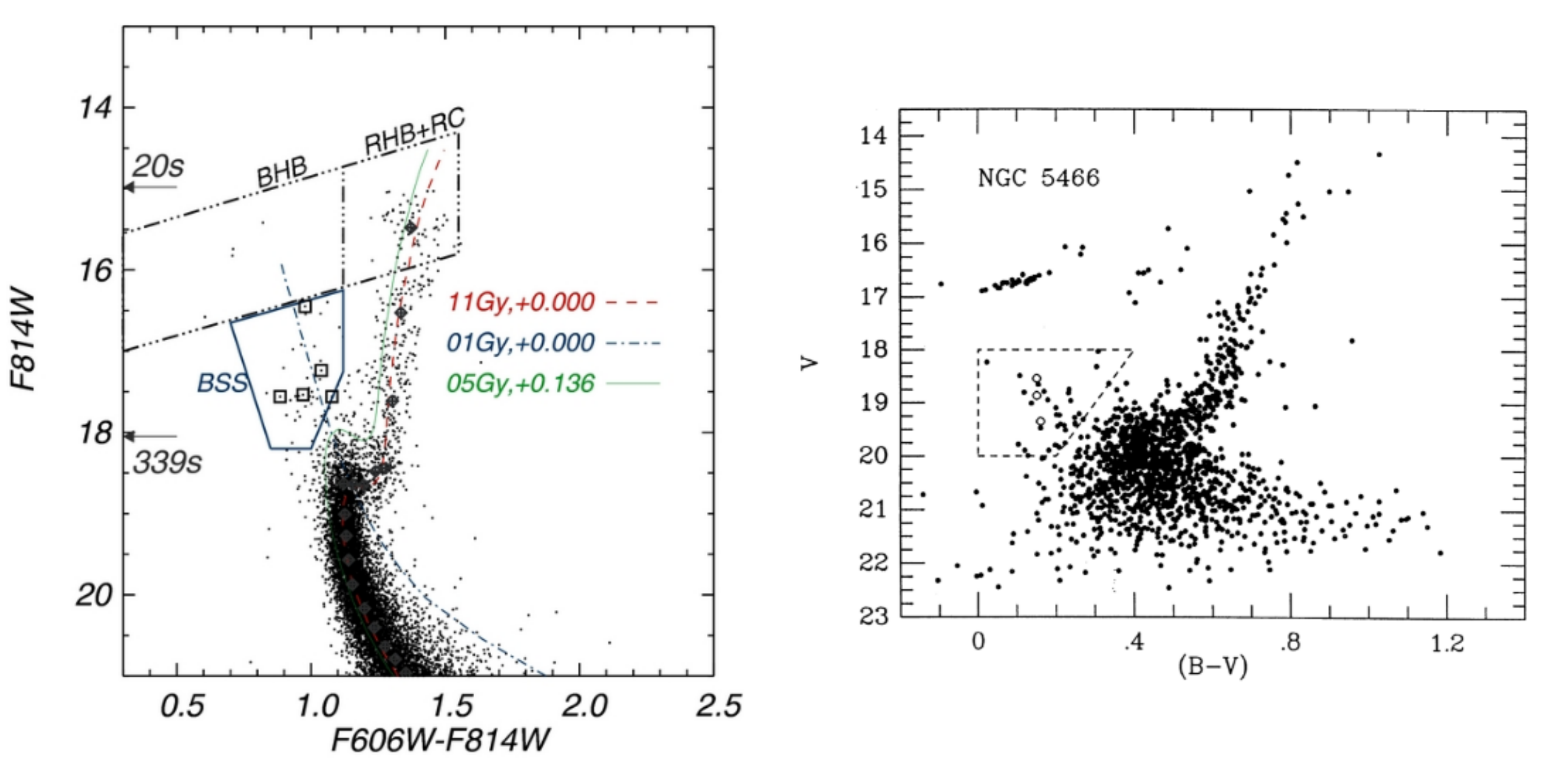}
%
% If no graphics program available, insert a blank space i.e. use
%\picplace{5cm}{2cm} % Give the correct figure height and width in cm
%
\caption{\emph{(left panel)} colour-magnitude diagram of the Galactic Bulge sample of \cite{cl11}. 
W UMa stars in the field are denoted by boxes; \emph{(right panel)} Colour-magnitude 
diagram of the GC NGC 5466 studied by \cite{ma90}.  W UMa stars are indicated by open circles.  These figures are reproduced 
with permission from the AAS.}
\label{prefig:18}       % Give a unique label
\end{figure}
\end{center}

Recently \cite{cl11} reported the discovery of BSSs in the Galactic Bulge.  They identify BSSs produced by mergers of W UMa stars\index{W UMa star} that 
follow the path of AW UMa\index{AW UMa} \cite{pa07} to FK Com\index{FK Comae star} \cite{bo81}.  They use these BSSs to place a conservative upper limit on 
the percentage of genuinely young stars in their sample.  Such young stars have interesting consequences for the history of 
the Bulge\index{Bulge}.  The strong resemblance of the Bulge observational material to that of \cite{ma90} for the GC NGC 5466\index{NGC 5466}, indicated by the side-by-side 
CMDs of NGC 5466 and the Bulge sample in Fig.~\ref{prefig:18}, invites a common explanation for at least a portion of the bulge BSSs.  

%%%%%%%%%%%%%%%%%%%%%%%%%%%%%%%%%%%%%%%%%% FIGURE 18
%%%%%%%%%%%%%%%%%%%%%%%%%%%%%%%%%%%%%%%%%% FIGURE 18

The ratios of variable to constant BSSs in the two CMDs, $5/30=0.167$ and $3/29=0.103$, are statistically indistinguishable.  From admittedly 
shaky estimates of BS and W UMa lifetimes, Mateo et al. \cite{ma90} concluded that all of the BSSs in NGC 5466 can be understood as mergers of W UMa stars.  
This conclusion, coupled with the great preponderance of long-period binaries in the distribution of Duquennoy \& Mayor \cite{du91}, provides a 
convenient rationale for the disparate specific frequencies of BSSs in GCs and the halo field.

Clarkson et al. \cite{cl11}, working with a much more complicated dataset, cautiously reach a suitably more modest conclusion, namely, that the Bulge BS population 
with $P<10$ d comprises at {\it most half} of all bulge BS binaries.  This is more or less consistent with their specific frequency for Bulge 
BSSs, $\sim1.23$, which is markedly smaller than that of the halo field.   If (almost) all the remainder are BSS binaries of longer period, an 
expectation based on halo experience, they arrive at an upper limit of  $\sim3\%$ for the young ($<5$Gyr) population of the bulge.  
Bensby et al.~\cite{ben12} argue for the existence of an intermediate-age population based on spectral analysis of Bulge dwarfs observed during 
microlensing\index{microlensing} events.  The age distribution of stars in the Bulge has arisen once again, like the Phoenix, as a new direction for research.

\subsection{Halo Streams}
\label{presubsec:4.3}

Morrison et al. \cite{mo00} included BMP stars in their list of tracers to investigate structure in the halo produced by satellite accretion, but I 
could find no subsequent results based on their use. Newberg, Yanny \& Willet \cite{ne09} identified a new polar-orbit stream (\emph{Cetus Polar Stream})  
spatially coincident with the Sagittarius\index{Sagittarius stream} trailing tidal trail,  but with systematically lower metallicity ([Fe/H]$=-2.1$), in which the specific 
frequency of BSSs differs from that in Sagittarius.  More recently Koposov et al. \cite{ko11} claim that the BSSs belong mainly to 
Sagittarius and the BHB belong mainly to the Cetus stream\index{Cetus stream}.  This conclusion requires gross, unlikely differences in specific 
frequencies of the two streams.  Streams in the halo\index{halo} is a fascinating but difficult subject in its infancy, so interpretation of results best 
awaits further investigation

\section{Metal-rich A-type Stars Above the Galactic Plane:  Another Inconvenient Truth}
\label{presec:5}

\begin{center}
\begin{figure}[h]
\sidecaption
\includegraphics[width=119mm]{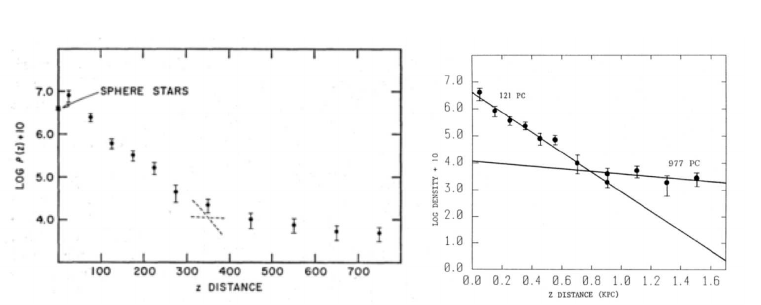}
%
% If no graphics program available, insert a blank space i.e. use
%\picplace{5cm}{2cm} % Give the correct figure height and width in cm
%
\caption{Space densities of early A-type stars near the North and South Galactic Poles versus distance from the Galactic plane 
according to: left panel, \cite{pe69}; right panel \cite{la88b}.  The figures are reproduced with permission from the AAS.}
\label{prefig:19}       % Give a unique label
\end{figure}
\end{center}

No review of FBSSs would be complete without a digression to the metal-rich A-type stars far above the Galactic plane\index{galactic plane}.  
The existence of such stars was first recognised by Perry \cite{pe69} who sought to determine 
the force law perpendicular to the Galactic plane by use of early A-type main sequence stars near the north Galactic pole, for which 
accurate distances could be derived from suitably calibrated Str\"omgren photometry\index{Str\"omgren photometry}.  Perry \cite{pe69} constructed a density distribution perpendicular to the plane, 
shown in the left panel of Fig.~\ref{prefig:19}, that produced an unbelievable force law.  

%%%%%%%%%%%%%%%%%%%%%%%%%%%%%%%%%%%%%%%%%% FIGURE 19
%%%%%%%%%%%%%%%%%%%%%%%%%%%%%%%%%%%%%%%%%% FIGURE 19

He was obliged to consider that the A-type stars more than 400 pc above the plane were a mixture of two populations.  This result prompted Rodgers
\cite{ro71} to investigate the A-type stars at the south Galactic pole (SGP) identified by \cite{ph68}.  He identified 21 
stars on $-0.04<(B-V)_o< 0.32$ with ``normal'' abundances of calcium and thought of no satisfactory explanation.  A decade later 
Rodgers, Harding  \& Sadler \cite{ro83} confirmed Rodgers \cite{ro71} first abundances, used their radial velocities and a believable force law 
(\cite{o60,in66}) to calculate $Z_{max}$ values that range from 1.5 to 8 kpc. They hypothesised that these stars were formed during a 
recent ($<10^9$yr ago) collision with a gas-rich satellite galaxy.

\begin{center}
\begin{figure}[h]
\sidecaption
\includegraphics[width=119mm]{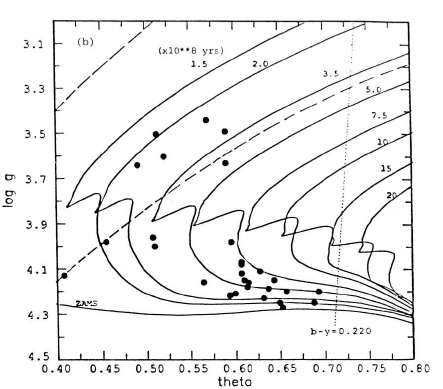}
%
% If no graphics program available, insert a blank space i.e. use
%\picplace{5cm}{2cm} % Give the correct figure height and width in cm
%
\caption{Locations of early A-type stars among RYI isochrones, according to \cite{la88b}.  This figure is 
reproduced with permission from the AAS.}
\label{prefig:20}       % Give a unique label
\end{figure}
\end{center}

Lance \cite{la88a,la88b} investigated this speculation anew by compiling a catalogue of 305 early type stars brighter than $V=15$ in 218 sq. deg at the SGP.  
From Str\"omgren photometry and the $byK$ equivalent of Pier's \cite{pi83} $BVK$ diagram she identified 29 Population I A-type stars which lie in $1<z(kpc)<11$
and $-0.6<$[Ca/H]$<0.2$, and have radial velocity dispersion $\sigma=62$ km $s^{-1}$.  The density distribution of her stars at the SGP shown 
in the right panel of Fig.~\ref{prefig:19}  bear the same two-slope signature of Perry's distribution in the left panel.  The locations of these 29 stars in the grid of 
revised Yale Isochrones (\cite{gr87}) translated to the (log $g, \Theta$) plane ($\Theta=5040/T_{eff}$) and displayed in Fig.~\ref{prefig:20}. 
indicate ages less than 0.6 Gyr.  

%%%%%%%%%%%%%%%%%%%%%%%%%%%%%%%%%%%%%%%%%% FIGURE 20
%%%%%%%%%%%%%%%%%%%%%%%%%%%%%%%%%%%%%%%%%% FIGURE 20

After considering various alternatives, Lance arrived at the earlier conclusion of Rodgers et al. \cite{ro83}: about 0.6 Gyr ago a slightly metal-poor, 
but still gas-rich, satellite merged with the disc of the Galaxy.  During the collision stars formed ``\emph{that do not partake of the usual 
age-abundance-kinematics relationships shown by other Galactic stellar groupings}'' (Lance \emph{verbatim}). 

So far as I am aware, Lance's work has never been refuted.  Rather, it simply has been ignored because, I suspect, her results do not conform to current dogma 
about Milky Way satellite\index{Milky Way satellite} encounters.  According to ADS her paper (\cite{la88b}) has received a total of 56 citations in 24 years, the last being 5 years ago in 
2007.  In contemporary vernacular, the astronomical community ``\emph{never lets facts stand in the way of a good idea}''.

\section{Abundance Issues}
\label{presec:6}

Three abundance topics are of particular importance for FBSSs: (a) Li\index{lithium} (and Be) as diagnostics of deep mixing in FBSSs and in 
the RGB/AGB antecedents of their present white dwarf companions (b) the $\alpha$-elements\index{$\alpha$-element} as possible discriminants for satellite accretion, 
and (c) neutron-capture elements\index{s-process element} as constraints on AGB models, thermohaline convection, and dilution.

\subsection{Lithium}
\label{presubsec:6.1}

Lithium deficiencies in blue stragglers of various ilks relative to those of the Spite plateau \cite{sp82} have been recognised 
for two decades \cite{pr91,ho91,gl94,ca05}.  More recently 
\cite{bo07} have added Be deficiencies to the problem.  Because $^7$Li is burned to $^4$He ash at temperatures greater 
than $\sim2\times10^6$ K,  Li deficits are widely believed to be consequences of mixing processes that are plausible consequences of collision\index{collision}, 
merger\index{merger}, or mass transfer\index{mass transfer} from a cool giant.  These mixers are by no means the only possibilities. Pinsonneault et al. \cite{pi99,pi02} consider 
rotation-driven circulation, while Ryan et al. \cite{ry02} discuss rotation produced by angular momentum\index{angular momentum} transfer during binary mass transfer.  
They suggest that such transfer could produce the rapid axial rotations (sic!), $v \sin i$ values of 8.3, 7.6, and 5.5 kms$^-1$, of three 
Li-deficient stars below the MSTO.  This suggestion can be reframed as an hypothesis amenable to observational test.  
Finally, in consideration of the ``Lithium Dip'' in young stellar populations~\cite{bo86}, it was, perhaps, inevitable that 
radiative levitation, gravitational settling \cite{mi91}, and associated mas -loss \cite{de97}
would be investigated.

The link between blue stragglers and the small fraction of Li-deficient stars below the MSTO, required by observation \cite{ma97}, is 
a natural consequence of power-law mass spectra adopted for the initial primaries and secondaries of primordial binaries used in the toy model of PS00.  
This link was also noted by Ryan et al. \cite{ry01,ry02}.  Worry by \cite{ry01,ry02} that some Li deficient MS stars are not known binaries will only 
become a real problem after these stars have been subjected to the same careful long-timescale scrutiny accorded the disc
\cite{du91}, halo \cite{la02}, and BMP (PS00) samples. Norris et al. \cite{no97} note that Li deficiency is not always accompanied 
by AGB\index{asymptotic giant branch star} carbon\index{carbon} and s-process enrichments.  For some unstated reason they do not consider the rational possibility that mass transfer from ordinary RGB 
\index{red giant} companions will deplete Li without affecting other [X/Fe].

Most recently Masseron et al. \cite{mas12} have made an extensive investigation of Li in CEMP-s stars.  Among the several important issues addressed in 
their paper is the unequivocal demonstration, provided by CS 22949-008, that mass transfer binaries populate the main sequence well below the MSTO.  
Equally interesting, they present a detailed discussion of the role of dilution, as parameterised by Gallino and his colleagues \cite{bi12} and 
modified by thermohaline mixing \cite{va94}, in creating the dispersion of Li abundance in the Spite plateau. Detailed discussion of their 
results is beyond the scope of this review; I only note that they \cite{mas12} provide yet another empirical argument for the notion that FBSSs represent 
an extension of the family of mass transfer binaries to locations above (hotter than) the MSTO.  Finally, Glebbeek et al. \cite{gl10} calculated surface abundances 
of blue straggler models and comment that it may be possible to use Li as a tracer to distinguish between formation by mass transfer and formation by collision.  

%\clearpage
%%%%%%%%%%%%%%%%%%%%%%%%%%%%%%%%%%%%%%%%%% FIGURE 21
\begin{center}
\begin{figure}[h]
\sidecaption
\includegraphics[width=119mm]{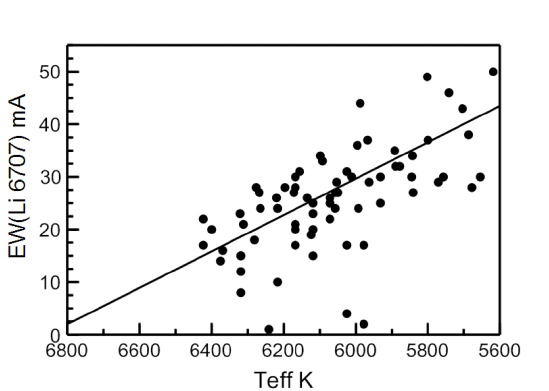}
%
% If no graphics program available, insert a blank space i.e. use
%\picplace{5cm}{2cm} % Give the correct figure height and width in cm
%
\caption{A plot of equivalent width of Li I (6707$\AA$) versus effective temperature for the main sequence stars studied by \cite{th94}.}
\label{prefig:21}       % Give a unique label
\end{figure}
\end{center}
%%%%%%%%%%%%%%%%%%%%%%%%%%%%%%%%%%%%%%%%%% FIGURE 21

One caution about detection of Li in FBSSs should be mentioned.  Due to increasing ionisation of Li I (ionisation potential$=5.4$eV) the 
strength of Li I $6707\AA$ decreases steadily at temperatures above MSTO values, as shown in Fig.~\ref{prefig:21}, a plot of equivalent width of  Li I 6707$\AA$ versus 
$T_{eff}$  along the Spite plateau made from data in Table 2 of \cite{th94}.  Linear extrapolation into the FBS region by use of the line drawn through the data in Fig.~\ref{prefig:21} indicates that equivalent widths for normal Spite-plateau abundances fall below 10 m$\AA$ at $T_{eff}>6500$ K.  Considering that most of the 
FBSSs found in the HK and HES surveys are fainter than $V=14$, detection of Li with the requisite resolution ($R>25000$) and signal-to-noise 
ratio (S/N$>50$) in most FBSSs can only be accomplished with large telescopes\index{telescope}.
 
\subsection{Alpha/Fe}
\label{presubsec:6.2}

The relatively lower values of [$\alpha$/Fe] found for red giants in Milky Way satellite galaxies (see Fig.~\ref{prefig:10} in section 2 above) suggests the possibility that 
[$\alpha$/Fe] might serve as a discriminant with which to distinguish FBSSs accreted from Milky Way satellites from FBSSs home-grown in the halo.  
The data for FBSSs, summarised in Table~\ref{pretab:1}, indicate that this might be so.  I sorted the data in Table 7 of PS00 first into two groups by metal 
abundance at [Fe/H]$=-1$ and then into binary and constant velocity groups within each abundance group.  I calculated average values of [$\alpha$/Fe] 
for these four groups, first treating Mg and Ca as the $\alpha$-elements\index{$\alpha$-element} in column 4 of Table 1, and then including Ti, frequently treated as an 
$\alpha$-element, in column 6.  

\begin{table}
\caption{[$\alpha$/Fe] behavior among FBSSs}
\label{pretab:1}       % Give a unique label
%
% Follow this input for your own table layout
%
%\begin{tabular}{p{2cm}p{2.4cm}p{2cm}p{4.9cm}}
%\begin{tabular}{p{1cm}p{1cm}p{2cm}p{2cm}p{2cm}p{2cm}p{2cm}p{2cm}}
\begin{tabular}{cccccccc}
\hline\noalign{\smallskip}
 &   &  &  $<[$MgCa/Fe$]>$   &  $<[$MgCa/Fe$]>$  &  $<[$MgCaTi/Fe$]>$   &  $<[$MgCaTi/Fe$]>$ & \\
 \hline
Population  & [Fe/H]  & [Fe/H] &  Avg.   &  SD(avg) & Avg.   &  SD(mean) & n \\
\hline
SB &$<-1$ & --1.69 & 0.42 & 0.03 & 0.44 & 0.03 & 23 \\
RVC &   &  &  &  &  &  &   \\
difference & & & 0.15 & 0.14 & & & \\
SB &$>-1$ & --0.45 & 0.24 & 0.07 & 0.25 & 0.06 & 12 \\
RVC & $>-1$  &--0.48 & 0.22 & 0.04  & 0.27  & 0.06 & 4  \\
difference & & & 0.02 &  &  --0.02 & &  \\
\noalign{\smallskip}\svhline\noalign{\smallskip}
\noalign{\smallskip}\hline\noalign{\smallskip}
\end{tabular}
\end{table}

The two cases are indistinguishable.  The last column of Table 1 contains the number of stars in each group. For the FBS 
stars with [Fe/H]$<-1$ the average [$\alpha$/Fe] is lower among the constantÐvelocity stars by 0.15 dex. The differences in columns 4 and 6 exceed 
the sums of the standard deviations of the means and are comparable to the amounts by which red giants of satellite galaxies lie below Galactic stars 
in Fig.~\ref{prefig:11}.  The differences found for the samples with disc metallicity are insignificant.  The abundances were derived from spectra of low-to moderate 
S/N for modest numbers of stars, so these results should be regarded as tentative. They are presented here as incentive for a definitive investigation.

Finally, we only mention that Ivans \cite{iv03}, on the basis of three stars, promote the existence of yet another population of metal-poor 
stars ([Fe/H]$\sim-2$) in the Galactic halo\index{halo}.  They are characterised by unusually low abundances of the $\alpha$-elements Mg, Si, and Ca and 
the neutron-capture elements Sr, Y, and Ba, and other peculiarities in abundances of the Fe-peak group.  One of the three 
stars, CS 22966-043, that define this putative new population is a pulsating binary blue straggler \cite{pre99}.

\subsection{The Neutron-Capture Elements}\index{s-process element}
\label{presubsec:6.3}

Bisterzo et al. \cite{bi12} have interpreted the abundances of 94 s-process enriched metal-poor C-stars, the CEMP-s stars\index{CEMP-s star}, by comparison with appropriate 
AGB models.  I confine my discussion below to 4 FBSSs among the 27 MS stars in their~\cite{bi12} sample.  To interpret the observed abundances 
of any CEMP-s star the following parameters must be specified: (a) chemical abundances, [X/Fe], in the envelopes, hence ejecta, of various AGB models, 
(b) a set of standard abundances for the companion star,  (c) the mass of ejecta captured by the companion, and (d) the envelope mass of the receiving 
star into which AGB ejecta are mixed.

If mixing were confined to the diminutive convective envelope, 0.001 $M/M_{\odot}$, of a typical FBSS with $T_{eff}\sim6400$ K (\cite{pi01}), 
the wind accretion of Boffin \& Jorissen \cite{bo88} would be more than adequate.  However, Stancliffe et al. \cite{st07} complicated the discussion by noticing that 
the enhanced molecular weight of overlying He- and C-enriched AGB accreta will induce thermohaline mixing\index{thermohaline mixing} throughout $\sim$90\% of the mass of a typical MS star.  
In a rejoinder Thompson et al. \cite{th08} argue that helium and the heavy elements settle in the surface layers of the receiver during the main-sequence 
lifetime ($\sim$3 Gyr) of a typical (1.3 $M_{\odot}$) AGB star employed by Bisterzo et al.~\cite{bi12}, thus creating a stabilising $\mu$Ðgradient\index{$\mu$ gradient} that confines 
thermohaline mixing to a thin surface layer until the stabilising $\mu$-gradient is flattened by sufficient accretion of high-$\mu$ matter.
The various issues of mixing associated with CEMP-s stars are reviewed by Stancliffe et al. \cite{st10}. 

Bisterzo et al. \cite{bi12} combine items (c) and (d) above into their dilution parameter, $dil=\log M_{CE}/M_{AGBacc}$, the log$_{10}$ of the ratio of mass 
of the convective envelope of the receiver to the mass of AGB accreta. They then try various combinations of AGB compositions and dilution factors 
to achieve the best agreement between observed and predicted abundances for each star.  
Details of the AGB models and element-by-element model-fitting procedures are beyond my competence. I restrict my discussion to aspects 
of their work pertinent to the FBSSs. 

Physical data for the four FBSSs studied by Bisterzo et al. \cite{bi12} are presented in Table 2.  Column 6 contains  $(B-V)_o$ values of the MSTO inferred from Y$^2$
isochrones\index{isochrone} for age =13.5 Gyr and the individual stellar [Fe/H] values in column 2 by use of  the curve in Fig.~\ref{prefig:13}.  
Column 7 contains the amount by which each star lies blueward of its MSTO.  Finally, column 8 contains the minimum accreted stellar mass inferred 
from the Y$^2$ isochrones. 

\begin{table}
\caption{FBSSs for which minimum accreted masses can be estimated}
\label{pretab:2}       % Give a unique label
%
% Follow this input for your own table layout
%
%\begin{tabular}{p{2cm}p{2.4cm}p{2cm}p{4.9cm}}
%\begin{tabular}{p{1cm}p{1cm}p{2cm}p{2cm}p{2cm}p{2cm}p{2cm}p{2cm}}
\begin{tabular}{cccccccc}
\hline\noalign{\smallskip}
%1 & 2  & 3 &  4    & 5  & 6  & 7  & 8\\
 %\aline
   &    &   &      &  star & TO   &  star-TO &  \\
\hline
star &	[Fe/H] &	$T_{eff}$ & log g &	$(B-V)_0$ & $(B-V)_0$ &	$d(B-V)_0$ & $min~M/M_{\odot}$ \\
CS 22887-048	&--1.70 &	6500 &	3.35 &	0.353 &	0.408 &	--0.055 &	0.09 \\
CS 29497-030	&--2.57 &	7000 &	4.10 &	0.296 &	0.364 &	--0.068 &	0.11 \\
CS 29526-110	& --2.06 &	6800 &	4.10	& 0.322 &	0.386 &	--0.064 &	0.11 \\
CS 29528-028 &	--2.86 &	6800 &	4.00 &	0.322 &	0.355 &	--0.033 &	0.06 \\
\noalign{\smallskip}\svhline\noalign{\smallskip}
\noalign{\smallskip}\hline\noalign{\smallskip}
\end{tabular}
\end{table}

They are minimum masses because there is no way to know the initial masses. These minimum masses can easily be 
provided by the wind accretion calculations of Boffin \& Jorissen~\cite{bo88}, and they are more than sufficient to completely fill the surface convection 
zones of these stars, but they require dilution parameters of order unity if the AGB material has been mixed into $\sim0.7~M/M_{\odot}$ of 
receiver envelope.  I include these calculations merely to indicate that the means are at hand to place additional observational constraints on the 
AGB models used to explain CEMP-s stars.

\subsection{How Many Evolved Mass-transfer Binaries Are There?}
\label{presubsec:6.4}

This topic was belabored by me earlier \cite{pre09}. To estimate the amount of mass transferred\index{mass transfer} requires a theory of accretion\index{accretion}. 
Boffin \& Jorissen \cite{bo88}, in a limited exploration of their parameter space, concluded that accretion in an AGB wind could produce a Ba star
like classic $\zeta$ Capricorni (\cite{bo80}).  Subsequently, Han et al.\cite{ha95} and Theuns, Boffin \& Jorissen \cite{th96} concurred that accretion from an 
AGB wind might produce Ba star\index{barium star} and CH star \index{CH star} companions in binaries with periods up to 100,000 d or more, a range that extends beyond the 
frequency maximum, near P$=50,000$ d, for F- and G-type dwarfs in the solar neighbourhood \cite{du91}.  On the other hand, 
Boffin \& Za\'cs \cite{bo94} observed that overabundances of s-process elements decline systematically with increasing orbital period (separation), 
as they must, in a recognisable albeit imperfect correlation.  Their results are in qualitative agreement with those of Jorissen \& van Eck\cite{jo00}
who found a steep drop in numbers of Ba star spectroscopic binaries with P$>3000$ d, the number falling to zero at 10,000 d.  
Failure to find such binaries is not a radial-velocity precision problem, but it may be an observer persistence problem. The $K_1$ value of a 0.8 M$_{\odot}$
star in a 10,000 d orbit with a 0.5 M$_{\odot}$ white dwarf companion is 4 kms$^{-1}$, easily detectable by all modern high-resolution spectrographs, 
if someone has the patience to monitor radial velocities on a 30-year timescale.  My gloss on all of this is that any census of CEMP-s stars 
is conditioned by observer persistence and the sensitivity of survey detection criteria.  Thresholds for detection skew the observed orbital 
period (size) distribution toward lower values and such bias acts to produce a discrepancy between the period distributions of 
Jorissen \& van Eck \cite{jo00} and Duquennoy \& Mayor \cite{du91}.

\section{Pulsating Blue Stragglers}\index{pulsating star}
\label{presec:7}

SX Phe stars\index{SX Phe star} are metal-poor pulsating blue stragglers found in GCs  and the halo field.  See Cohen \& Sarajedini \cite{co12} and references 
contained therein. The issues of SX Phe stars are complex and multifarious.  I confine my attention here to only two of them.

\subsection{Non-variable Stars in the SX Phe Instability Strip}
\label{presubsec:7.1}

PS00 found one large-amplitude SX Phe star and two probable low-amplitude versions \cite{pre98, pre99} among 40 BMP binaries.  
They noted that other low-amplitude pulsators might contribute to the radial velocity residuals of some of their binary orbits, but it is clear 
from examination of velocity residuals in Fig. 7 of PS00 that velocity amplitudes of putative pulsators in many of these binaries would have 
to be near or below the 1 kms$^{-1}$ level. {\it High-amplitude pulsators like SX Phe are a distinct minority among FBS spectroscopic binaries.}

\begin{center}
\begin{figure}[h]
\sidecaption
\includegraphics[width=119mm]{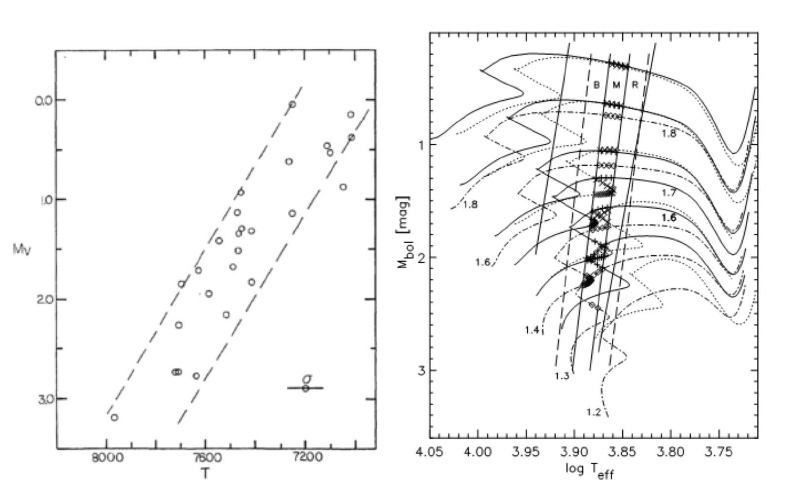}
%
% If no graphics program available, insert a blank space i.e. use
%\picplace{5cm}{2cm} % Give the correct figure height and width in cm
%
\caption{\emph{(left panel)} The instability strip for $\delta$ Scuti and SX Phe stars from \cite{mcn97}.  This figure is reproduced with permission 
of Publications of the Astronomical Society of the Pacific; \emph{(right panel)} McNamara's empirical strip, indicated by ``M'', in the theoretical 
instability strip marked by heavy solid lines, according to \cite{pe99}.  This figure is reproduced with permission 
of Astronomy \& Astrophysics.}
\label{prefig:22}       % Give a unique label
\end{figure}
\end{center}

\emph{Kepler}\index{Kepler satellite} data for the $\delta$ Scuti stars\index{$\delta$ Scuti star}, Population I analogues of SX Phe stars, discussed by Balona \& Dziembowski~\cite{ba11}, pre-empt 
all previous photometric inquires into this topic.  In spite of exquisite photometric precision that permits detection of very 
low-amplitude ($<100$ ppm) variables, they find that the $\delta$ Scuti fraction of stars in the instability strip barely reaches
{\it 0.5 only in the narrow temperature interval} $7500<T_{eff}$ (K)$<8000$, and falls to very low levels below 7000 K and above 8500 K.  
{\it The great majority of FBSSs in the $\delta$ Sct instability strip are non-variable even by stringent Kepler standards}, a circumstance contrary 
to all previous experience with the instability strips of the RR Lyrae stars and classical Cepheids. This concentration to a narrow interval 
of $T_{eff}$, less than 300 K, is borne out approximately by the locations of high amplitude $\delta$ Scuti and SX Phe stars in the ($M_V$, $T_{eff}$)
plane of McNamara \cite{mcn97} reproduced here in the left panel of Fig.~\ref{prefig:22}.

%%%%%%%%%%%%%%%%%%%%%%%%%%%%%%%%%%%%%%%%%% FIGURE 22
%%%%%%%%%%%%%%%%%%%%%%%%%%%%%%%%%%%%%%%%%% FIGURE 22

McNamara's narrow \emph{High Amplitude Delta Scuti} (HADS) strip, identified 
by the letter ``M'' in the right panel of Fig.~\ref{prefig:22}, cannot be predicted theoretically according to Petersen \& Christensen-Dalsgaard \cite{pe99}.  
Furthermore, stability analysis \cite{pa00} does not explain the paucity of SX Phe stars within the much broader temperature 
boundaries (thick solid lines in the right panel of Fig.~\ref{prefig:22}).  

SX Phe stars identified in the Fornax dSph galaxy\index{Fornax} \cite{po08} are not confined to the HADS strip of Figure~\ref{prefig:22} above, but the stars 
are faint and photometric errors blur this conclusion.  As in the \emph{Kepler} field the SX Phe stars comprise only a small fraction of the total BS population 
of Fornax. Balona \& Dziembowski \cite{ba11} suggest that unspecified damping mechanisms operate to suppress pulsation in many BSSs.  
According to them, ``\emph{Perhaps we need to wait for the development of non-linear, non-radial calculations to properly address this problem.}''
Explanations for many SX Phe phenomena are still in disarray.

\subsection{RRLYR-02792, Archetype of a New Kind of Mass-transfer Pulsator}
\label{presubsec:7.2}

Pietrzy{\'n}ski et al. \cite{pi12} describe a new kind of pulsating star found in an OGLE eclipsing binary\index{eclipsing binary}, RRLYR-02792 
\cite{so11}.  Its light and radial velocity variations resemble, approximately, those of an RRb star\index{RR Lyrae star} with period $P=0.627$ d, but 
its dynamical mass, 0.26 M$_{\odot}$, is incompatible with a location on the horizontal branch\index{horizontal branch star}. Pietrzy{\'n}ski et al. \cite{pi12} believe that they have 
caught an Algol-type binary\index{Algol system} in a short-lived evolutionary phase following mass exchange\index{mass transfer}, in which the pulsator possesses a partially 
degenerate helium core and a small H-shell-burning envelope.  The pulsator is now shrinking as it evolves toward the hot subdwarf\index{subdwarf} region 
at a rate compatible with the decline of its pulsation period, $8.4-2.6\times10^{-6}$ d yr$^{-1}$.  Wonders never cease. 

\section{Odds and Ends}
\label{presec:8}
\subsection{Mass Transfer in Hierarchical Triples}
\label{presubsec:8.1}

Approximately half of the short-period ($P<30$ d) binaries in the solar neighbourhood reside in multiple systems \cite{to06}, so 
we expect to find them among main sequence CEMP-s stars\index{CEMP-s star}, whether above or below the MSTO, if the statistical properties of disc and 
halo binary populations are similar. Apparent confirmation of such expectation is provided by CS 22964-161 \cite{th08} and 
CS 22949-008 \cite{mas12}, double-lined binaries\index{double-lined spectroscopic binary} in which both components are carbon and s-process enriched.  Both components 
of CS 22964-161 have subgiant characteristics in contrast to CS 22949-008, in which both components lie well below the MSTO.  In both 
cases the AGB donor must have been the remote third star of an hierarchical triple. This circumstance puts a lower limit on the semi-major 
axis of the AGB orbit, hence upper limits on the density of the AGB wind at the close binary and on the accretion of AGB gas by that binary system.   

Soker \cite{so04} initiated inquiry into the very case of interest here, wind accretion\index{wind accretion} by the close binary companions of a remote AGB in an 
hierarchical triple system\index{triple system}.  The problem is extremely complex, requiring specification of the inclination angle of orbital planes and treatment 
anew of Bondi \& Hoyle~\cite{bo44} accretion in an AGB wind, including assumptions about cooling times and angular momentum transfer\index{angular momentum transfer} in the 
accretion column.   With regard to the latter Davies \& Pringle \cite{da80} called attention to a conceptual problem for angular momentum transfer in 
an inhomogeneous medium that was not resolved in subsequent investigations by Livio et al. \cite{li86a,li86b}   Accretion rates for the two binary 
components may differ, in which case they will possess different surface enrichments by the AGB accreta. Investigation of accretion in hierarchical 
triples adds a new dimension to McCrea's initial ideas.

\subsection{Heresy}
\label{presubsec:8.2}

I conclude my review of issues that attend FBS and mass transfer, by brief enumeration of a few mildly disquieting observational facts.  

Lithium\index{lithium} again.  After building a case for AGB enrichment of the surface layers of CS 22964-161, enrichment in gas that has endured deep mixing in cool 
giant atmosphere, Thompson et al. \cite{th08} were obliged to appeal to some process by which lithium is produced in the AGB star prior to mass transfer, and 
by just the right amount that elevates it to the Spite plateau.  Extraordinary coincidence, one may suppose.  The process must be a bit less than 
extraordinary, because it must also account for the detection of appreciable lithium in two additional relatively bright (nearby) MS 
CEMP-s stars, CS 22898-027 \cite{th92} and LP 706-7 \cite{no97}.

\begin{figure}[h]
\sidecaption
\includegraphics[width=119mm]{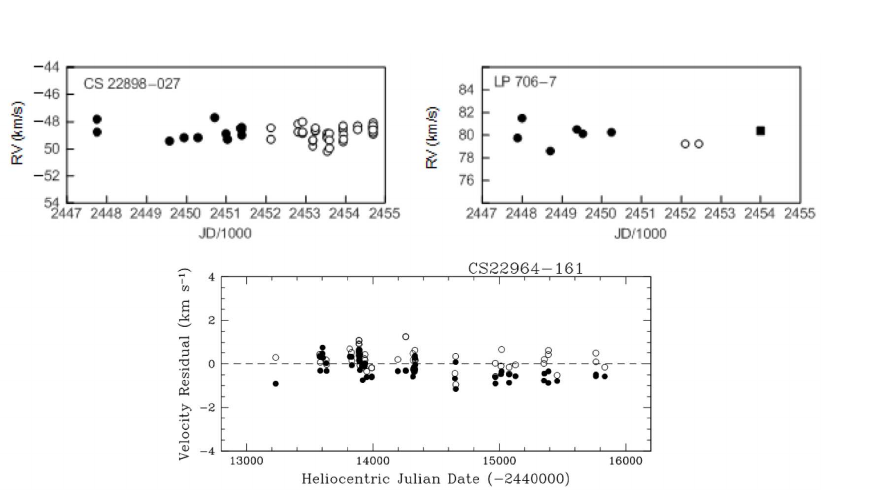}
%
% If no graphics program available, insert a blank space i.e. use
%\picplace{5cm}{2cm} % Give the correct figure height and width in cm
%
\caption{\emph{(top left)} Radial velocity versus Julian Date for MS CEMP-s star CS 22898-027; \emph{(top right)} radial; velocity versus Julian Date for the MS CEMP 
star LP 706-7.  The figures are reproduced with permission of Publications of the Astronomical Society of Australia; \emph{(bottom)} residuals of 
individual radial velocity observations of the primary (solid circles) and secondary (open circles) components of the SB2 CEMP-s system 
CS 22964-161 (Ian Thompson, private communication).}
\label{prefig:23}       % Give a unique label
\end{figure}

The three aforementioned stars share another property illustrated in Fig.~\ref{prefig:23}: extant observations provide no clear evidence for orbital motion.  
The results for CS 22898-027 and LP 706-7 are taken from \cite{pre09};  those for CD 22964-161 are velocity residuals with respect to 
predictions of the binary orbit under continuing investigation by Ian Thompson (private communication).  The centre of mass velocity ought 
to vary sooner or later, if the third AGB relic is present.

%\clearpage 
%%%%%%%%%%%%%%%%%%%%%%%%%%%%%%%%%%%%%%%%%% FIGURE 23
%%%%%%%%%%%%%%%%%%%%%%%%%%%%%%%%%%%%%%%%%% FIGURE 23

CS 22880-074 \cite{pre09} and TY Gru\index{TY Grucis} (\cite{pre11} are additional examples of apparently misbehaving CEMP-s stars.  I leave it to the reader 
to consider whether the radial velocity behaviours of these stars (never mind Li!) are all statistical flukes, chance consequences of small orbital 
inclinations, or are they warning signals about the universal AGB paradigm for production of s-process elements.

\input{refpreston}